\documentclass[conference,letterpaper]{IEEEtran}
\usepackage{times,amsmath}
\usepackage{graphics}
\usepackage{graphicx}
\usepackage{xspace}
\usepackage{subfigure}
\usepackage{amsmath}
\usepackage{amssymb}
\usepackage{array}
\usepackage{weiwAlgorithm}
\usepackage{url}
\usepackage{pdfpages}
\usepackage{color}

\newskip\subfigtoppskip \subfigtopskip = -1mm
\newskip\subfigcapskip \subfigcapskip = -1mm

\renewcommand{\baselinestretch}{0.941}

\usepackage[
normalsections, % Don't modify section headers.
normalfloats,  % Don't modify float parameters.
normalindent,  % Don't modify paragraph indentation.
normaltitle,   % Don't modify the formatting of document title.
normalleading, % Don't modify interline spacing.
normalmargins, normalbib, normalbibnotes % Note: DO NOT comment these arguments
]{savetrees}
\newtheorem{theorem}{Theorem}

\newcommand{\pefp}{\kwnospace{PEFP}\xspace}

\newcommand{\kwnospace}[1]{{\ensuremath {\mathsf{#1}}}}

\newcommand{\ei}{\end{itemize}}
\newcommand{\ee}{\end{enumerate}}

\newcommand{\beqn}{\begin{eqnarray*}}
\newcommand{\eeqn}{\end{eqnarray*}}

\newcounter{ccc}

\newcommand{\eat}[1]{}

\def\subfigcapskip{2pt}
\long\def\comment#1{}

\makeatletter

\newcommand{\Rmnum}[1]{\expandafter\@slowromancap\romannumeral #1@}
\makeatother

\begin{document}

\title{PEFP: Efficient $k$-hop Constrained $s$-$t$ Simple Path Enumeration on FPGA}

\author{{Zhengmin Lai$^{\dagger}$, You Peng$^{\S}$, Shiyu Yang$^{\dagger}$, Xuemin Lin$^{\S}$, Wenjie Zhang$^{\S}$} %

\vspace{1.6mm}\\
\fontsize{10}{10}
\selectfont\itshape
$^\dagger$East China Normal University; $^\S$The University of New South Wales\\
\fontsize{9}{9} \selectfont\ttfamily\upshape
zmlai@stu.ecnu.edu.cn;
you.peng@unsw.edu.au;\\
syyang@sei.ecnu.edu.cn;
\{lxue, zhangw\}@cse.unsw.edu.au \\}

\maketitle
\begin{abstract}
Graph plays a vital role in representing entities and their relationships in a variety of fields, such as e-commerce networks, social networks and biological networks. Given two vertices $s$ and $t$, one of the fundamental problems in graph databases is to investigate the relationships between $s$ and $t$. A well-studied problem in such area is $k$-hop constrained $s$-$t$ simple path enumeration. Nevertheless, all existing algorithms targeting this problem follow the DFS-based paradigm, which cannot scale up well. Moreover, using hardware devices like FPGA to accelerate graph computation has become popular. Motivated by this, in this paper, we propose the first FPGA-based algorithm \pefp to solve the problem of $k$-hop constrained $s$-$t$ simple path enumeration efficiently. On the host side, we propose a preprocessing algorithm Pre-BFS to reduce the graph size and search space. On the FPGA side in \pefp, we propose a novel DFS-based batching technique to save on-chip memory efficiently. In addition, we also propose caching techniques to cache necessary data in BRAM, which overcome the latency bottleneck brought by the read/write operations from/to FPGA DRAM. Finally, we propose a data separation technique to enable dataflow optimization for the path verification module; hence the sub-stages in that module can be executed in parallel. Comprehensive experiments show that \pefp outperforms the state-of-the-art algorithm JOIN by more than 1 order of magnitude by average, and up to 2 orders of magnitude in terms of preprocessing time, query processing time and total time, respectively.
\end{abstract}

\section{Introduction}
\label{sec:introduction}

Graph is a ubiquitous structure modeling entities and their relationships in various areas like e-commerce networks, social networks and biological networks~\cite{peng2018efficient, DBLP:journals/vldb/LiuYLQZZ20, peng2020answering, lai2019distributed}. One of the fundamental and important problems in graph databases is $k$-hop constrained $s$-$t$ path enumeration~\cite{qin2019towards, DBLP:journals/pvldb/QiuCQPZLZ18}; that is, given a directed and unlabelled graph $G$, source node $s$, target node $t$, and hop constraint $k$, we aim to enumerate all $s$-$t$ paths such that the number of hops of each path is no more than $k$. In this paper, same as many existing studies on this problem, we only consider the \textit{simple} path (i.e., a path with no repeated nodes) since a path containing cycles is less interesting and may significantly increase the total number of $s$-$t$ paths. Note that for presentation simplicity, we abbreviate $s$-$t$ $k$-hop constrained simple path as $s$-$t$ $k$-path in this paper.

During $s$-$t$ $k$-path computation, we have to frequently access neighbors of vertices in the graph. Real-life graphs usually follow power-law random distribution; that is, most vertices have a small degree, while some have a large degree~\cite{chung2003eigenvalues}. However, modern CPUs are not an ideal way to deal with such data accesses: they do not offer high parallelism, and their caches do not work effectively for irregular graph processing that has little or no temporal and spatial locality. GPUs, on the other hand, offer massive parallelism, but the performance can be significantly affected when the internal cores do not execute the same instruction (i.e., warp divergence), which is common in graphs with varying degrees~\cite{besta2019graph}.

FPGA has shown its substantial advantages over multi-core CPU in terms of parallelism~\cite{FPGA}. For instance, one FPGA card can easily parallelize a loop with 1,000 iterations, while we have to find a host equipped with 1,000 CPU cores to offer the same parallelism. In addition, compared with GPU, FPGA is more energy-efficient, and can handle irregular graph processing with more stable parallelism by fully exploiting its pipeline mechanism~\cite{besta2019graph}. Therefore, in this paper, we reconsider the problem of $s$-$t$ $k$-path enumeration on FPGA.

\vspace{1mm}
\noindent \textbf{Applications.}
We introduce the applications of $s$-$t$ $k$-path enumeration on FPGA as follows. 

\begin{itemize}
    \item \textit{E-commerce Networks.} 
    A cycle in e-commerce networks indicates that there might exist fraudulent activities among the participants~\cite{ICWCMMC2007}. 
    To detect such activities, Alibaba Group has developed a system in~\cite{DBLP:journals/pvldb/QiuCQPZLZ18}; that is, when a new transaction is submitted from account $t$ to account $s$, the system will perform $s$-$t$ $k$-path enumeration to report all newly produced cycles. 
    Since response time is very critical to the fraud detection system, it is necessary to speed up the $s$-$t$ $k$-path queries in e-commerce networks.
    In this paper, we choose FPGA due to its parallel and re-programmable properties.
    %Note that without imposing hop constraint $k$, the system will be overwhelmed by the tremendous reported results with false alarms.
    
    \vspace{0.2mm}
    \item \textit{Social Networks.} For two users $s$ and $t$ in a social network, we may wonder to what extent $t$ is influenced by or similar with $s$~\cite{kimura2006tractable}. One can achieve this by enumerating all the simple paths from $s$ to $t$ with hop constraint $k$. As querying the $s$-$t$ $k$-paths in a vast social network is very time-consuming, it is essential to accelerate such queries using FPGA.
    
    \vspace{0.2mm}
    \item \textit{Biological Networks.} It is known that $s$-$t$ $k$-path enumeration is one of the most important pathway queries in biological networks~\cite{article_leser}; that is, given two substances $s$ and $t$, one can figure out the chains of interactions from $s$ to $t$ by enumerating all paths from $s$ to $t$ with hop constraint $k$. As biological networks are quite sensitive to the response time of pathway queries, it is necessary to accelerate $s$-$t$ $k$-path queries through FPGA.
\end{itemize}

We have to emphasize that, besides the hop constraint, we can for sure impose other constraints to $s$-$t$ path queries. For instance, one can apply label constraints to the vertices in social networks such that only specific types of users will be considered. Note that although we study the problem of $s$-$t$ $k$-path enumeration in unlabelled graphs in this paper, our solutions can be easily extended to solve it in labelled graphs; that is, we can deal with the label constraints in preprocessing stage to filter out the vertices and edges that satisfy the constraints.

% \vspace{1mm}
% \noindent \textbf{Existing Solutions.}
% T-DFS~\cite{DBLP:conf/iwoca/RizziSS14} and T-DFS2~\cite{grossi2018efficient} are the first two algorithms solving the problem of $s$-$t$ $k$-path enumeration. Given an intermediate path $p$ and its one-hop successor $u$, both T-DFS and T-DFS2 aggressively check the shortest distance $l$ from $u$ to $t$ without touching any vertices in $p$, which is the art of ``\textit{never fall in the trap}". If $len(p) + 1 + l > k$ holds, $u$ will be pruned by $p$, where $len(p)$ denotes the length of $p$ and $len(p) = |p| - 1$. However, the two algorithms show poor performance in practice due to the expensive verification cost~\cite{qin2019towards}. 

% Qiu et al. study the problem of $s$-$t$ simple cycle detection~\cite{DBLP:journals/pvldb/QiuCQPZLZ18}, in which HP-Index is proposed. The idea of HP-Index is to index paths among ``hot-points" to accelerate path queries, where a hot-point is a vertex whose degree is larger than the predefined threshold $\delta$. Most recently, Peng et al. have proposed their competitive algorithm JOIN in \cite{qin2019towards}. Observing the expensive verification cost of T-DFS and T-DFS2, JOIN is designed following the art of ``\textit{never fall in the same trap twice}". As a result, JOIN significantly outperforms the other algorithms\cite{qin2019towards}.

\vspace{1mm}
\noindent \textbf{Challenges.} We present the challenges of solving the problem of $s$-$t$ $k$-path enumeration on FPGA as follows.
\begin{itemize}
    \item \textit{Exponential Search Space and Expensive Verification Cost.} The main challenge of $s$-$t$ $k$-path enumeration is the huge search space even if $k$ is very small, because the number of results grows exponentially w.r.t $k$. Moreover, the tremendous number of intermediate results incurs expensive cost for path verification, which ensures that there are no repeated vertices along the path. It is inefficient in both response time and memory usage to simply enumerate all $s$-$t$ $k$-paths with duplicate vertices, and then verify them.
    
    \item \textit{Non-trivial Implementation on FPGA.} Due to the huge intermediate results using BFS-based framework, all existing solutions follow the DFS-based paradigm for better performance~\cite{qin2019towards, DBLP:journals/pvldb/QiuCQPZLZ18, DBLP:conf/iwoca/RizziSS14, grossi2018efficient}. However, DFS-based algorithms cannot be pipelined on FPGA because of the data dependencies among iterations. Thus existing algorithms cannot be straightforwardly implemented on FPGA. In addition, since CPU usually has an order of higher frequency than FPGA, it requires careful design on FPGA to achieve better performance than CPU.
    
    \item \textit{Limited FPGA on-chip Memory.} Although BFS-based algorithms can be pipelined on FPGA, there is very limited FPGA on-chip memory (BRAM); hence, we have to frequently transfer intermediate results between BRAM and FPGA's external memory (DRAM) when using BFS-based paradigm, which significantly affects the overall performance. Therefore, one of the biggest challenges of solving this problem comes from how to deal with the huge intermediate data on FPGA efficiently to achieve good performance.
\end{itemize}

Consequently, it is rather challenging to design an efficient $s$-$t$ $k$-path enumeration algorithm on FPGA that tames both computational hardness and on-chip memory bottleneck.

\vspace{1mm}
\noindent \textbf{Contributions.}
Our contributions in this paper are summarized as follows:
\begin{itemize}
    \item To the best of our knowledge, none of the existing $s$-$t$ $k$-path enumeration algorithms can be directly adapted to the FPGA side. Therefore, we are the first to propose an efficient algorithm to solve this challenging problem on FPGA.
    
    \item On the host side, we develop a preprocessing algorithm \textbf{Pre-BFS} that can not only greatly reduce the search space in finding $s$-$t$ $k$-paths, but also can finish in satisfactory time. 
    
    \item On the FPGA side, we design an efficient algorithm \textbf{PEFP}. In \pefp, we first propose a novel DFS-based batching technique \textit{Batch-DFS} to overcome the FPGA on-chip memory bottleneck. Then we further propose \textit{caching} techniques to improve the read/write latency by reducing memory accesses to DRAM. Finally, we propose a \textit{data separation} technique to fully parallelize the path verification module.
    
    \item We conduct comprehensive experiments on 12 real datasets to demonstrate the superior performance of our proposed algorithm \pefp compared with the state-of-the-art algorithm JOIN, where \pefp runs on the Xilinx Alveo U200 FPGA card~\footnote{https://www.xilinx.com/products/boards-and-kits/alveo/u200.html}. More specifically, the experimental results show that \pefp outperforms JOIN by more than 1 order of magnitude by average, and up to 2 orders of magnitude in terms of preprocessing time, query processing time and total time, respectively.
\end{itemize}

\vspace{1mm}
\noindent \textbf{Roadmap.}
The rest of the paper is organized as follows. Section~\ref{sec:related} surveys important related works. Section~\ref{sec:preliminary} gives the formal definition of the problem studied in this paper, and introduces the existing solutions. Section~\ref{sec:framework} presents the overall framework. We then propose our software preprocessing algorithm in Section~\ref{sec:soft_preprocessing} and hardware implementation details in Section~\ref{sec:hardwareImpl}. Extensive experiments are conducted in Section~\ref{sec:experiment}. Finally, Section~\ref{sec:conclusion} concludes the paper.
\section{Related Work}
\label{sec:related}

In this section, we review closely related works.

\subsection{Simple Path Enumeration and Counting}
\label{subsec:path}

There are many existing works studying the problem of enumerating $s$-$t$ simple paths (e.g.,~\cite{bohmova2018computing,bookknuth11,DBLP:conf/ambn/YasudaSM17}). However, what they focus on is how to construct a succinct presentation of these simple paths, thus we can efficiently enumerate the simple paths without explicitly storing each path. Note that their algorithms are not competitive for the problem of $s$-$t$ simple path enumeration, and can only handle small graphs with thousands of vertices. Birmele \emph{et. al} studied the problem of $s$-$t$ simple path enumeration in~\cite{DBLP:conf/soda/BirmeleFGMPRS13}, but the solution they proposed can only handle undirected graphs.

% counting version
The counting of $s$-$t$ simple paths is also a well-known $\#P$ hard problem, which has been extensively studied with different approaches such as recursive expressions of an adjacency matrix (e.g.,~\cite{DBLP:journals/ipl/Golovko72,DBLP:journals/corr/GiscardKW16}).
However, their counting approaches cannot be extended to efficiently enumerate hop-constrained simple paths in a trivial manner without materializing the paths during the computation, which will easily blow up the main memory even for a small $k$.

\subsection{Shortest Path Enumeration}
\label{subsec:topk_SP}

Given two vertices $s$ and $t$ in a graph, the end-to-end shortest path computation from $s$ to $t$ is one of the most important graph queries. In addition to the classical $s$-$t$ shortest path computation, there are several variants where a set of paths are considered.

The problem of top-$k'$ shortest paths has been intensively studied in the literatures (e.g.,~\cite{gotthilf2009improved, DBLP:journals/Yen71}). To solve $s$-$t$ $k$-path enumeration problem, we can keep on invoking the top-$k'$ shortest simple path algorithm by increasing $k'$ until the shortest path detected exceeds the distance threshold $k$. However, this naive method is not competitive because we have to enforce the output order of the paths according to their distances. 

A considerable number of literatures have been published studying the constrained shortest path problem recently (e.g.,~\cite{rivera2016mathematical,shi2017multi}). This problem can be defined as finding the shortest path between two vertices on a network whenever the traversal of any arc/vertex consumes certain resources. The problem of diversified shortest path has been intensively studied in the literatures as well (e.g.,~\cite{liu2017finding,talarico2015k}), which consider both distance and diversity of $s$-$t$ shortest paths. However, due to their focus on identifying the shortest paths, they cannot be adapted to $s$-$t$ $k$-path enumeration in a trivial manner.

\subsection{Shortest Path Computation on FPGA}
\label{subsec:graph-pro-in-fpga}
Recently there emerge many literatures aiming at accelerating shortest path computation on FPGA (see~\cite{besta2019graph} for a survey). Tommiska et al.~\cite{tommiska2001dijkstra} implemented single source shortest path (SSSP) algorithm on FPGA using adjacency matrix stored in BRAM, which limits its graph size that can be handled. Unlike the previous approach, Zhou et al.~\cite{zhou2015accelerating} solved SSSP with the graph stored in DRAM and the algorithm is fully pipelined. Bondhugula et al.~\cite{bondhugula2006parallel} proposed to solve all-pairs-shortest-paths (APSP) problem on FPGA, which is to find shortest path between all pairs of vertices in the graph. The graph is stored in DRAM and only when the required slices of graph are streamed to BRAM. Betkaoui et al.~\cite{betkaoui2012parallel} studied APSP for unweighted graphs by running BFS from each vertex, and its key idea for optimizing memory accesses is to use BRAM for random memory accesses and use DRAM for sequential accesses. Nevertheless, none of these algorithms can be directly adapted to $s$-$t$ $k$-path enumeration problem because they can only identify the shortest paths rather than enumerate all $s$-$t$ $k$-paths.
\section{Preliminary}
\label{sec:preliminary}

In this section, we first give the formal definition of $s$-$t$ $k$-hop constrained simple path enumeration problem, then we present a brief introduction to the existing solutions, namely T-DFS~\cite{DBLP:conf/iwoca/RizziSS14}, T-DFS2~\cite{grossi2018efficient}, HP-Index~\cite{DBLP:journals/pvldb/QiuCQPZLZ18}, and JOIN~\cite{qin2019towards}. We summarize important notations in TABLE~\ref{tb:notations}.

\begin{table}
    \centering
    \begin{tabular}{|p{2cm}|p{5cm}|}
      \hline
      \textbf{Notation}   & \textbf{Definition}             
      \\ \hline \hline
      $G, G_{rev}$        & a graph, its reverse graph
      \\ \hline
      $V, E$              & graph vertex set, edge set
      \\ \hline
      $p, v \rightsquigarrow v'$ & a path, a path from $v$ to $v'$ \\ \hline
      $s$, $t$, $k$       & source and target vertex, hop constraint 
      \\ \hline
      $s$-$t$ $k$-path    & $k$-hop constrained path from $s$ to $t$
      \\ \hline
      $len(p)$            & length of path $p$, where $len(p) = |p| - 1$
      \\ \hline
      $sd(v, v')$         & shortest distance from $v$ to $v'$      
      \\ \hline
      $sd(v, v'|p)$       & shortest distance from $v$ to $v'$ without touching any vertex in $V(p)$
      \\ \hline
      $bar[u]$            & shortest distance from $u$ to $t$       \\ \hline
      $\mathcal{P}$       & path set of buffer area in BRAM     
      \\ \hline
      $\mathcal{P'}$      &  path set of processing area in BRAM   \\ \hline
      $\mathcal{P_D}$     & path set in DRAM
      \\ \hline
      $\mathcal{S}[i]$            & one hop successors of path $\mathcal{P'}[i]$
      \\ \hline
    \end{tabular}
%\vspace{-2mm}
\caption{Summary of Notations}
\label{tb:notations}
\vspace{-2mm}
\end{table}

\subsection{Problem Definition}
\label{subsec:prob_def}
A directed graph $G$ is represented as $G=(V, E)$, where $V(G)$ is the vertex set of $G$, and $E(G) \subseteq V(G) \times V(G)$ is the directed edge set of $G$. If the context is clear, we use \textbf{successor} or \textbf{neighbor} to refer ``\textit{out-going neighbor}". Let $G_{rev}$ denote the reverse graph of $G$, where $V(G_{rev}) = V(G)$ and for each edge $(v_1, v_2) \in E(G)$, there is a corresponding edge $(v_2, v_1) \in E(G_{rev})$. We say $G'$ is an induced subgraph of $G$ if $V(G') \subseteq V(G)$ and $E(G') = \{(v_1, v_2) | (v_1, v_2) \in E(G), v_1 \in V(G') \wedge v_2 \in V(G')\}$. A path $p$ from vertex $v$ to vertex $v'$ is a sequence of vertices $v = v_0, v_1, ..., v_n = v'$ such that for each $i \in [0, n - 1]$, $(v_i, v_{i+1}) \in E(G)$. We use $p(v, v')$ or $p(v \rightsquigarrow v')$ to denote a path from $v$ to $v'$. A simple path is a loop-free path that contains no duplicate nodes. We use \textbf{path} to refer ``\textit{simple path}" if the context is clear. The length of a path $p$ is denoted as $len(p)$, where $len(p) = |p| - 1$. We say path $p$ is a $k$-hop constrained path if it satisfies $len(p) \leq k$. Given two vertices $v, v' \in V(G)$ and a path $p$, we use $sd(v, v')$ to denote the shortest distance from $v$ to $v'$, and use $sd(v, v'|p)$ to denote the shortest distance from $v$ to $v'$ without touching any vertex in $V(p)$. We say $u$ is a successor of path $p$ if $u$ is an out-going neighbor of the last vertex in $p$.

\vspace{1mm}
\noindent \textbf{Problem Statement.}
In this paper, we study the FPGA-based $k$-hop constrained $s$-$t$ simple path enumeration problem. Specifically, given a directed and unlabelled graph $G$, source vertex $s$, target vertex $t$, and hop constraint $k$, we use $\mathcal{R}$ to represent the paths such that $\mathcal{R} = \{p$ $|$ $len(p) \leq k$, $p$ is a simple path and $p$ starts with $s$ and ends with $t\}$. We target developing FPGA-based algorithms to efficiently enumerate all paths in $\mathcal{R}$.

\begin{figure}
\vspace{-0.2cm}
\centering
\includegraphics[width=0.9\columnwidth]{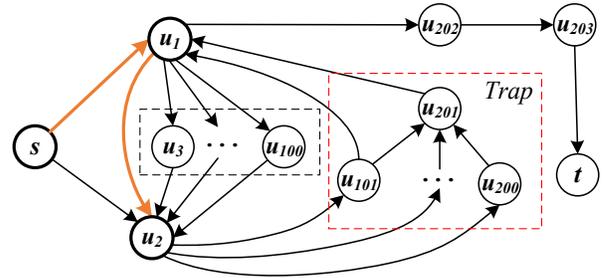}
\vspace{-0.2cm}
\caption{Key Idea of BC-DFS with $k = 7$}
\label{fig:bc-dfs}
\end{figure}

\subsection{Existing Solutions}
\label{subsec:existing-solutions}
\vspace{1mm}
\noindent \textbf{T-DFS and T-DFS2.} In~\cite{DBLP:conf/iwoca/RizziSS14}, T-DFS is proposed to solve $s$-$t$ $k$-path enumeration problem in directed graphs. T-DFS carefully explores the out-going neighbors of a vertex and ensures that every search branch comes up with at least one $s$-$t$ $k$-path, which is the art of ``\textit{never fall in the trap}". Specifically, given a current path $p$, T-DFS aggressively computes the shortest distance $sd(u, t|p)$ for each successor $u$ of $p$, and $u$ will not be explored if $len(p) + 1 + sd(u, t|p) > k$. T-DFS2~\cite{grossi2018efficient} follows the same aggressive verification strategy as T-DFS, while it can reduce shortest path distance computation by skipping some vertices associated with only one output in the following search. Nevertheless, T-DFS and T-DFS2 show poor performance in practice due to the expensive verification cost~\cite{qin2019towards}.

\vspace{1mm}
\noindent \textbf{HP-Index.} In~\cite{DBLP:journals/pvldb/QiuCQPZLZ18}, a novel indexing technique HP-Index is proposed to continuously maintain the pairwise paths among hot points (i.e., vertices with high degree). Enumerating all $s$-$t$ $k$-paths in HP-Index can be concluded as follows: (1) Perform DFS from $s$ with search depth at most $k$, and record the path and backtrack when encountering a hot point; (2) Perform a reverse DFS from $t$ in the same way; (3) Find the indexed paths among the hot points involved in the above computation; (4) Concatenate the paths from steps (1), (2) and (3) to identify $s$-$t$ $k$-paths. It is reported in~\cite{qin2019towards} that HP-Index can only achieve good performance on the extremely skewed graph dataset which has relatively small number of $s$-$t$ $k$-paths.

\vspace{1mm}
\noindent \textbf{JOIN.} Peng et al.~\cite{qin2019towards} proposed the state-of-the-art algorithm JOIN to enumerate all $s$-$t$ $k$-paths in a directed and unlabelled graph. JOIN designs an efficient pruning technique BC-DFS motivated by ``\textit{never fall in the same trap twice through learning from mistakes}". The idea of BC-DFS is shown in Fig.~\ref{fig:bc-dfs}, where the hop constraint $k$ is set to 7. We regard the current path $p$ as a stack $S$. In Fig.~\ref{fig:bc-dfs}, node $s, u_1$ and $u_2$ have been pushed into $S$. After finishing DFS with $S$, we know that there is no valid $s$-$t$ $k$-path w.r.t $S$. Then BC-DFS will set $u_2.bar = k + 1 - len(S)$, which is $6$ in this example. When $u_2$ is unstacked and we push $u_3$ into the stack, it will not fall in the same ``trap" like $u_2$ did before because it will check if $len(S) + 1 + u_2.bar \leq k$ holds. In this example, $len(S) + 1 + u_2.bar = 2 + 1 + 6 = 9 > 7$, hence $u_2$ will be pruned by $u_3$, ..., $u_{100}$ when $s$ and $u_1$ are in the stack, which significantly reduces the search space.

Given a path $p = (u_1, ..., u_n)$, its middle vertex is the $\lceil 
\frac{n}{2}\rceil$-th vertex. To avoid duplicate search, JOIN follows ``\textit{joining two paths}" framework by exploiting the middle vertices of $s$-$t$ $k$-paths. Its procedures can be concluded as follows: (1) Compute all middle vertices of $s$-$t$ $k$-paths, denoted by $\mathcal{M}$; (2) Add a virtual vertex $t'$ and put an edge $(u, t')$ for each $u \in \mathcal{M}$; (3) Compute $s$-$t'$ $(\lceil\frac{k}{2}\rceil + 1)$-paths $P_l$ using BC-DFS; (4) Add a virtual vertex $s'$ and put an edge $(s', u)$ for each $u \in \mathcal{M}$; (5) Compute $s'$-$t$ $(\lfloor\frac{k}{2}\rfloor + 1)$-paths $P_r$ using BC-DFS; (6) Join $P_l$ and $P_r$ to obtain the final results, where the join key is the node $u \in \mathcal{M}$, and a result path is valid iff it is a simple path and $u$ is its middle vertex. Consequently, JOIN outperforms all existing algorithms, namely T-DFS~\cite{DBLP:conf/iwoca/RizziSS14}, T-DFS2~\cite{grossi2018efficient}, and HP-Index~\cite{DBLP:journals/pvldb/QiuCQPZLZ18}.
\section{Framework Overview}
\label{sec:framework}

\begin{figure}
\vspace{-0.2cm}
\centering
\includegraphics[width=0.9\columnwidth]{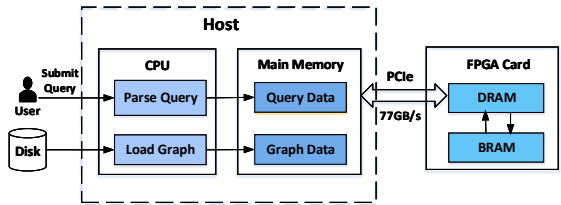}
\vspace{-0.2cm}
\caption{Overall Framework of CPU-FPGA System}
\label{fig:fram}
\end{figure}

In this section, we present our system's overall framework to solve the problem of $s$-$t$ $k$-path enumeration. The overview of the system architecture is illustrated in Fig.~\ref{fig:fram}. The workflow of the system can be concluded as follows:
\begin{enumerate}
    \item On the host side, the user first specifies the graph file, then the host loads the corresponding graph data and stores it in main memory. Once the graph loading process is finished, the host is ready to handle path queries submitted by user;
    
    \item When a new query comes in, the host parses the query to extract $s$, $t$ and $k$;
    
    \item The host starts its preprocessing  computation (for further details see Section~\ref{sec:soft_preprocessing}) to prepare the necessary data that will be transferred to FPGA DRAM.
    
    \item Based on the preprocessing, the host transfers the prepared data to FPGA DRAM through PCIe bus in DMA mode;
    
    \item Once all the input data arrive at DRAM, FPGA can start its computation to find all valid $s$-$t$ $k$-paths and return them to the host side through PCIe. The computation details on FPGA will be introduced in Section \ref{sec:hardwareImpl}.
    
\end{enumerate}
\section{Software Preprocessing}
\label{sec:soft_preprocessing}
In this section, we present the preprocessing details on the host side. We first give a brief introduction to the preprocessing technique of the state-of-the-art algorithm \textbf{JOIN}. Then we propose our optimized preprocessing technique to further reduce the search space for future path expansion and verification.

Preprocessing aims to compute and prepare necessary data for a given algorithm. As for preprocessing in JOIN, a $k$-hop BFS is first conducted from source vertex $s$ to compute $sd(s, u)$ on $G$. Similarly, JOIN computes $sd(u, t)$ by conducting $k$-hop BFS from $t$ on $G_{rev}$. The shortest distance of those vertices that have not been touched during BFS is set to $k+1$. After finishing preprocessing, JOIN can start $s$-$t$ $k$-path computation, which is introduced in subsection~\ref{subsec:existing-solutions}.

Having investigated the preprocessing idea of JOIN, we find it can be optimized based on the following Theorem.

\begin{figure}
  \centering
    \subfigure[Graph $G$]{
    \label{fig:pre-obs-1}
    \includegraphics[width=0.47\columnwidth]{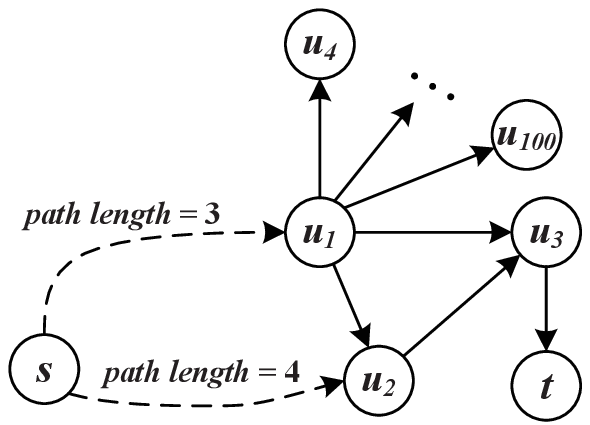}}
  \hfill
  \subfigure[Induced Subgraph after Removing Invalid Nodes with $k=5$]{
    \label{fig:pre-obs-2}
    \includegraphics[width=0.47\columnwidth]{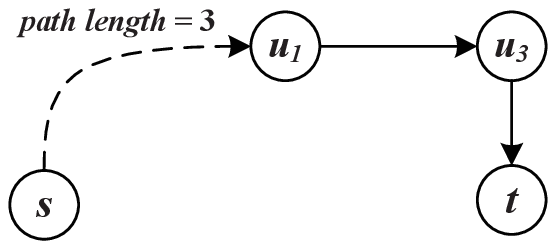}}
    \vspace{-0.2cm}
  \caption{Example of Preprocessing}
  \label{fig:preprocessing} %% label for entire figure
\end{figure}

\begin{theorem}
Given shortest distance maps $sd_s$ and $sd_t$, where $sd_s[v] = sd(s, v)$ and $sd_t[v] = sd(v, t)$ for any $v \in V(G)$, performing $s$-$t$ $k$-path enumeration on original graph $G$ is equivalent to doing it on subgraph $G' \subseteq G$ , where $G'$ is induced by the node set $\mathcal{N} \subseteq V(G)$ such that for each $u \in \mathcal{N}$, $sd_s[u] + sd_t[u] \leq k$.
\end{theorem}
\noindent 
\begin{proof}
Suppose there exists a valid $s$-$t$ $k$-path $p$ that contains a node $u$, where $u$ is not in $G'$. Thus we have $sd_s[u] + sd_t[u] > k$, which contradicts the premise $len(p) \leq k$. Hence the theorem holds.
\end{proof}

As illustrated in Fig.~\ref{fig:pre-obs-1}, when we set $k=5$, $5$-hop BFS from $s$ will include the whole graph $G$. Although $u_4$, ..., $u_{100}$ cannot reach $t$, they will still be visited in JOIN during BC-DFS, which results in useless search. After we remove the invalid nodes such that $sd_s[u] + sd_t[u] > 5$, the induced subgraph is shown in Fig.~\ref{fig:pre-obs-2}, where $u_2$, $u_4$, ..., $u_{100}$ and their corresponding edges are removed, hence the search space is greatly reduced.

Next, we show that a $(k-1)$-hop bidirectional BFS on $G$ is enough for preprocessing as follows.

We call a vertex $u$ is valid in preprocessing iff $sd_s[u] + sd_t[u] \leq k$. We use $G_{k-1}$ and $G_k$ to denote the induced subgraph after running $(k-1)$-hop and $k$-hop bidirectional BFS on $G$, respectively. We show that $(k-1)$-hop bidirectional BFS is enough by proving $V(G_{k-1})$ contains all valid vertices.

\begin{proof}
Suppose there is a valid vertex $u$ such that $u \in V(G_k)$ and $u \notin V(G_{k-1})$. Let $V'$ denote the vertices that $k$-hop bidirectional BFS can touch while $(k-1)$-hop cannot. It is obvious that $V'=\{u | sd_s[u] = k$ or $sd_t[u] = k \}$. Nevertheless, if $u$ is a valid node and $sd_s[u] = k$, then we have $u = t$, which contradicts the premise that $u \notin V(G_{k-1})$ because $t \in V(G_{k-1})$. Similarly, if $u$ is a valid node and $sd_t[u] = k$, then we have $u = s$, which contradicts the premise that $u \notin V(G_{k-1})$ because $s \in V(G_{k-1})$. Thus, $V(G_{k-1})$ includes all valid vertices.
\end{proof}

Based on the above observataions, we propose our preprocessing algorithm \textbf{Pre-BFS} which only needs to do $(k-1)$-hop bidirectional BFS to obtain the induced subgraph $G'$ as follows.
\begin{enumerate}
    \item Perform $(k-1)$-hop BFS from $s$ on $G$ to compute $sd_s$.
    \item Perform $(k-1)$-hop BFS from $t$ on $G_{rev}$ to compute $sd_t$;
    \item For each node $u \in sd_s \cap sd_t$, if $sd_s[u] + sd_t[u] \leq k$ holds, then we put it into node set $\mathcal{N}$.
    \item Return the subgraph $G'$ induced by $\mathcal{N}$ in $G$.
\end{enumerate}

When the preprocessing procedure is finished, we will send $s$, $t$, $sd_t$ and $G'$ to FPGA DRAM, where $G'$ is stored using ``Compressed Sparse Row" (CSR) format~\cite{CSR}. Note that we call $sd_t$ as barrier (denoted as $bar$) in the rest of the paper.
\section{Hardware Implementation}
\label{sec:hardwareImpl}

In this section, we first introduce our proposed algorithm \textbf{PEFP} to solve $s$-$t$ $k$-hop constrained \textbf{P}ath \textbf{E}numeration on \textbf{FP}GA. Then we present several optimizations to improve the performance of \pefp by fully utilizing the characteristics of FPGA.

\subsection{PEFP}
\label{subsec:pefp}

\pefp adopts the BFS-based paradigm, because BFS naturally enjoys great parallelism such as performing concurrent expansion for some intermediate results in a certain round. Therefore, we can easily apply pipeline optimizations to BFS-based algorithms to fully utilize the parallelism of FPGA.

In general, \pefp follows the \textbf{expansion-and-verification} framework, which can be dissected into three steps: (1) Expand the intermediate paths with one-hop successor vertices; (2) Verify if each expanded path is a valid path; (3) Write back the valid paths to the intermediate path set. The algorithm terminates when the intermediate path set is empty.

The details of \pefp are shown in Algorithm \ref{alg:pefp}. Given source vertex $s$, target vertex $t$, hop constraint $k$, barrier $bar$ and graph $G$, the algorithm computes and outputs all $s$-$t$ $k$-paths. We first initialize intermediate path set $\mathcal{P}$, $\mathcal{P'}$, $\mathcal{P}_D$, and one-hop successors set $\mathcal{S}$ with empty set (Line 1), where $\mathcal{P'}$ is a batch of paths fetched from $\mathcal{P}$, $\mathcal{P}_D$ represents the intermediate path set in DRAM, and $\mathcal{S}[i]$ denotes the one-hop successor vertex set of the $i$-th path in $\mathcal{P'}$ (denoted as $\mathcal{P'}[i]$). Then we push a path into $\mathcal{P'}$ consisting of just one vertex $s$ (Line 2). If $\mathcal{P'}$ is not empty, for each path $\mathcal{P'}[i] \in \mathcal{P'}$, we get its one-hop successors and put them into $\mathcal{S}[i]$ (Line 3-5). Next, we verify each successor $nbr \in \mathcal{S}[i]$ for every $\mathcal{P'}[i] \in \mathcal{P'}$ using Algorithm~\ref{alg:verify} (Line 6-9). If $nbr$ is a valid successor for $\mathcal{P'}[i]$, a new intermediate path $p$ will be generated and put into $\mathcal{P}$ by concatenating $nbr$ to $\mathcal{P'}[i]$ (Line 10-12). Note that when the size of $\mathcal{P}$ reaches our predefined threshold, we will flush $\mathcal{P}$ to DRAM to avoid BRAM overflow (Line 13-14). After all paths in $\mathcal{P'}$ have been processed in this batch, we fetch next batch of paths into $\mathcal{P'}$ using Algorithm \ref{alg:next_batch} (Line 15). \pefp terminates when $\mathcal{P'}$ is empty (Line 3).

\vspace{1mm}
Given an intermediate path $p$, its one-hop successor $u$, target vertex $t$, hop constraint $k$ and its barrier $b_u$, the verification module of checking whether $u$ is a valid successor for $p$ includes three stages, which is shown in Algorithm \ref{alg:verify}: (1) The first stage is \textbf{target check}. If $u$ equals to target vertex $t$, then we output a new result path $p'$ by concatenating $u$ to $p$ and return $false$ (Line 1-4); (2) The second stage is \textbf{barrier check}. If $len(p) + 1 + b_u > k$, then we return \textit{false} as it does not satisfy the hop constraint (Line 5-6); (3) The third stage is \textbf{visited check}. If $u$ has already appeared in $p$, then we return $false$ (Line 7-8). If $u$ passes the validity check of the three stages, we can say $u$ is a valid successor of $p$ (Line 9).

\begin{algorithm}[htb]
\SetVline % save space
\SetFuncSty{textsf}
\SetArgSty{textsf}
\small
\caption{\textbf{PEFP}($s$, $t$, $k$, $bar$, $G$)}
\label{alg:pefp}
\Input
{
	$s:$ source vertex \\
    $t:$ target vertex \\
    $k:$ hop constraint \\
    $bar:$ barrier array \\
    $G:$ graph \\
}
\Output{All $s$-$t$ $k$-paths}
\State{$\mathcal{P} \leftarrow \emptyset$; $\mathcal{P'} \leftarrow \emptyset$; $\mathcal{P}_D \leftarrow \emptyset$; $\mathcal{S} \leftarrow \emptyset$}
\State{$\mathcal{P'}.push(\{s\})$}
\While{$\mathcal{P'} \not= \emptyset$}
{
    \ForAll{$\mathcal{P'}[i] \in \mathcal{P'}$}
    {
        \State{$\mathcal{S}[i] \leftarrow$ one-hop successor vertices of $\mathcal{P'}[i]$ in $G$}
    }
    \ForAll{$\mathcal{S}[i] \in \mathcal{S}$}
    {
        \ForAll{$\mathcal{S}[i][j] \in \mathcal{S}[i]$}
        {
            \State{$nbr \leftarrow \mathcal{S}[i][j]$}
            \State{$isValid \leftarrow$ \textbf{Verify}($\mathcal{P'}[i]$, $nbr$, $t$, $k$, $bar[nbr]$)}
            \If{$isValid == true$}
            {
                \State{$p \leftarrow P'[i].push(nbr)$}
                \State{$\mathcal{P}.push(p)$}
            }
            \If{$\mathcal{P}$ is full}
            {
                \State{Flush $\mathcal{P}$ to $\mathcal{P}_D$}
            }
        }
    }
    \State{$\mathcal{P'} \leftarrow $ \textbf{NextBatch}$(\mathcal{P}, \mathcal{P_D})$ }
}
\end{algorithm}

\begin{algorithm}[htb]
\SetVline % save space
\SetFuncSty{textsf}
\SetArgSty{textsf}
\small
\caption{\textbf{Verify}($p$, $u$, $t$, $k$, $b_u$)}
\label{alg:verify}
\Input
{
	$p:$ path $p$ \\
	$u$: a successor of path $p$ \\
    $t:$ target vertex \\
    $k:$ hop constraint \\
    $b_u:$ barrier of vertex $u$ \\
}
\Output{$isValid:$ if $u$ is a valid successor of $p$}
\If(\tcc*[f]{Target Check}){$u == t$}
{
    \State{$p' \leftarrow p.push(u)$}
    \State{\textbf{output} $p'$}
    \State{\Return $false$}
}
\If(\tcc*[f]{Barrier Check}){$len(p) + 1 + b_u > k$}
{
     \State{\Return $false$}
}
\If(\tcc*[f]{Visited Check}){$u$ is contained in path $p$}
{
     \State{\Return $false$}
}
\State{\Return $true$}   
\end{algorithm}

\begin{algorithm}[htb]
\SetVline % save space
\SetFuncSty{textsf}
\SetArgSty{textsf}
\small
\caption{\textbf{NextBatch}($\mathcal{P}$, $\mathcal{P}_{D}$)}
\label{alg:next_batch}
\Input
{
	$\mathcal{P}$: intermediate path set in BRAM \\
    $\mathcal{P}_{D}:$ intermediate path set in DRAM \\
}
\Output{$\mathcal{P'}:$ a batch of intermediate paths}
\State{$\mathcal{P'} \leftarrow \emptyset$}
\State{$\Theta_1 \leftarrow $ batch size threshold of $\mathcal{P}_D$}
\State{$\Theta_2 \leftarrow $ batch size threshold of $\mathcal{P}$}
\If{$\mathcal{P} \not= \emptyset$}
{
    \State{$\mathcal{P'} \leftarrow$ \textbf{Batch-DFS}($\mathcal{P}$, $\Theta_2$)}
}
\Else
{
    \If{$\mathcal{P}_{D} \not= \emptyset$}
    {
         \State{$\mathcal{P} \leftarrow$ fetch a batch of paths from $\mathcal{P}_{D}$ with $\Theta_1$}
         \State{$\mathcal{P'} \leftarrow$ \textbf{Batch-DFS}($\mathcal{P}$, $\Theta_2$)}
    }
} 
\State{\Return $\mathcal{P'}$}
\end{algorithm}

\vspace{1mm}
\noindent \textbf{Correctness.} In Section \ref{sec:soft_preprocessing}, we have correctly calculated barrier data $bar$ for each vertex in the induced subgraph $G$ (recall that $bar[u] = sd(u, t)$). The expansion of $\mathcal{P'}$ starts with the path only containing vertex $s$. Therefore, the correctness of \pefp holds before the loop starts. For each iteration of expansion, given path $\mathcal{P'}[i]$ and its one-hop successor $nbr$, there are three cases to check in total: (1) Whether vertex $nbr$ is the target vertex; (2) Whether the path $\mathcal{P'}[i]$ exceeds the hop constraint when concatenating $nbr$ to $\mathcal{P'}[i]$; (3) Whether vertex $nbr$ has already appeared in $\mathcal{P'}[i]$. Only when $nbr$ passes all the three cases can we generate a new intermediate path $p = \mathcal{P'}[i].push(nbr)$. Therefore, we will not prune any valid paths during the verification and the correctness of each iteration holds. The algorithm terminates when $\mathcal{P'}$ is empty, suggesting that all of the intermediate paths have been processed. Hence the correctness of \pefp holds.

\subsection{DFS-based Batch Processing with Caching}
\label{subsec:batch_proc}

Intuitively, BFS-based path enumeration needs to store all intermediate results, causing notorious memory overhead to our system. To solve this challenging issue, we adopt the \textit{buffer-and-batch} technique. The general idea of buffer-and-batch aims to store huge intermediate paths in FPGA's external memory (DRAM), then read and process the data from DRAM by batch to avoid BRAM overflow.

Nevertheless, there exists another concern. Although DRAM capacity is much larger than BRAM, the read latency of DRAM takes 7-8 clock cycles while the read latency of BRAM is only 1 clock cycle. Based on that observation, we propose a caching-based technique to efficiently reduce the read/write operations from/to DRAM, thereby lowering the system latency.

\vspace{1.0mm}
\noindent \textbf{($\mathbf{1}$) Caching Intermediate Paths.} Targeting maximizing FPGA on-chip memory usage and minimizing the number of direct accesses to DRAM, we design two areas in BRAM, namely \textbf{buffer area} and \textbf{processing area}. As shown in Algorithm \ref{alg:next_batch}, the input $\mathcal{P}$ denotes the intermediate path set in BRAM, which is the \textbf{buffer area}; the input $\mathcal{P}_{D}$ denotes the intermediate path set in DRAM, which is the \textbf{external memory area}; the output $\mathcal{P'}$ represents a batch of intermediate paths we need to process next, which is called \textbf{processing area}. We first check the buffer area $\mathcal{P}$. If $\mathcal{P}$ is not empty, we fetch a batch of paths directly from $\mathcal{P}$ into $\mathcal{P'}$ (Line 4-5) using Algorithm \ref{alg:batch-dfs}. Note that we use $\Theta_2$ to denote the batch size threshold of $\mathcal{P}$ (Line 3), which is the capacity of $\mathcal{P'}$. Otherwise, we check $\mathcal{P}_{D}$. If $\mathcal{P}_{D}$ is not empty, we first fetch a batch of paths from $\mathcal{P}_{D}$ into $\mathcal{P}$ with batch size $\Theta_1$ (which is defined in Line 2), then fetch a batch of paths from $\mathcal{P}$ into $\mathcal{P'}$ using Algorithm \ref{alg:batch-dfs}, finally return $\mathcal{P'}$ (Line 6-10). Note that when we fetch a batch of paths from $\mathcal{P}_D$, we simply fetch from its tail with size $\Theta_1$ to avoid memory fragmentation; we do the same for the write operation. Thanks to the buffer area, we only need to read/write intermediate paths from/to DRAM when the buffer area is empty/full. By caching intermediate paths in BRAM, we can significantly reduce the data transfer between BRAM and DRAM, hence the overall performance is improved.

\vspace{1.0mm}
\noindent \textbf{($\mathbf{2}$) Caching Data Graph and Barrier.} From Algorithm \ref{alg:pefp} we know that we need to frequently access barrier data and get one-hop successors from data graph $G$ for a given path $p$. Learning from the merits by caching, we also cache the data graph and barrier information in BRAM. More specifically, we have pre-allocated three fixed-size arrays $vertex\_arr$, $edge\_arr$ and $bar\_arr$ to store vertex data, edge data and barrier data, respectively. When initializing the three arrays, we put as much data as possible into them from DRAM. When we access vertex, edge or barrier data, we always check the local BRAM array first instead of directly fetching it from DRAM. Thanks to the preprocessing to extract induced subgraph, we find that in most cases, we can fit the whole subgraph and barrier data in BRAM.

\begin{algorithm}[htb]
\SetVline % save space
\SetFuncSty{textsf}
\SetArgSty{textsf}
\small
\caption{\textbf{Batch-DFS}($\mathcal{P}$, $\Theta$)}
\label{alg:batch-dfs}
\Input
{
	$\mathcal{P}:$ intermediate path set \\
    $\Theta:$ batch size threshold \\
}
\Output{$\mathcal{P'}:$ a batch of intermediate paths}
\State{$\mathcal{P'} \leftarrow \emptyset$}
\State{$cnt \leftarrow 0$}
\State{$i \leftarrow$ index of the last path in $\mathcal{P}$}
\While{$i \not= 0$ }
{
    \State{$ptr_1, ptr_2 \leftarrow$ the end neighbor pointer of $\mathcal{P}[i]$}
   \State{$ptr_{last} \leftarrow$ the last neighbor pointer of $\mathcal{P}[i]$}
   \If{$ptr_1 + \Theta - cnt < ptr_{last}$}
    {
       \State{$ptr_2 \leftarrow ptr_1 + \Theta - cnt$}
    }         
    \Else{
        \State{$ptr_2 \leftarrow p_{last}$}
    }
    \State{$\mathcal{P'}.push(\mathcal{P}[i])$ with $ptr_1$ and $ptr_2$}
    \State{update start and end neighbor pointer of $\mathcal{P}[i]$ with $ptr_1$ and $ptr_2$}  
    \State{$cnt \leftarrow cnt + ptr_2 - ptr_1$}

    \If{$cnt < \Theta$}
    {
        \State{$i \leftarrow i - 1$}
    }
    \Else
    {
        \State{\textbf{break}}   
    }
}
\State{\Return $\mathcal{P'}$}
\end{algorithm}

\vspace{1.0mm}
\noindent \textbf{($\mathbf{3}$) Batch-DFS.} As mentioned before, only when the buffer area $\mathcal{P}$ is full will it do write operations to DRAM. Therefore, it is essential to design an efficient batching technique to save the memory in buffer area. One alternative is to change the batch size dynamically. For instance, one can reduce the batch size according to the free space in $\mathcal{P}$. Nevertheless, when there 
is only little space left in $\mathcal{P}$, the batch size will be set to a rather small value. In this case, most space in the processing area is wasted, which leads to a low level parallelism of \pefp. Thus this naive technique is inefficient.

It is challenging to design an efficient batching algorithm that both fully utilizes the space in \textbf{processing area} and saves the memory in \textbf{buffer area}. To overcome this challenge, we propose a novel DFS-based batching algorithm \textbf{Batch-DFS}, which is shown in Algorithm~\ref{alg:batch-dfs}. The motivation of this algorithm comes from the following observation.

\textit{Observation 1}: Given two paths $p_1$ and $p_2$ with $len(p_1) < len(p_2)$, suppose $p_1$ and $p_2$ have same number of successors, then $p_2$ will have a greater chance generating fewer intermediate paths than $p_1$ during one-hop expansion.

\begin{figure}
\vspace{-0.2cm}
\centering
\includegraphics[width=0.8\columnwidth]{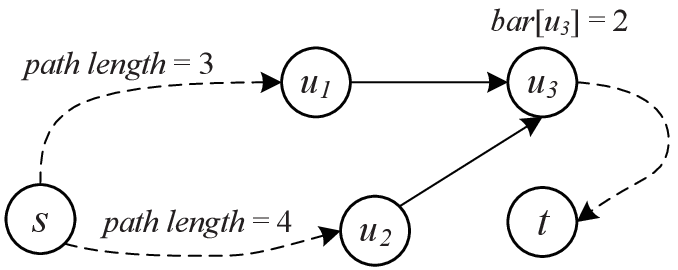}
\vspace{-0.2cm}
\caption{An Example for Processing Longer Path First with $k = 6$}
\label{fig:Obs-DFS}
\end{figure}

The observation is illustrated in Fig. \ref{fig:Obs-DFS}. Suppose the hop constraint $k$ is $6$, $p_1 = s \rightsquigarrow u_1$, $p_2 = s \rightsquigarrow u_2$, $len(p_1) = 3$, $len(p_2) = 4$, and path $p_1$ and $p_2$ have a same successor $u_3$, where $bar[u_3] = 2$. Clearly, $u_3$ will be pruned by $p_2$ because $len(p_2) + 1 + bar[u_3] > 6$, while it will not be pruned by $p_1$. Accordingly, when we process $p_2$ prior to $p_1$, it tends to produce fewer intermediate paths for $p_2$'s pruning power is stronger than $p_1$ in the barrier check stage. Specifically, when we process a path $p$ with $len(p) = k-1$, it will generate $0$ intermediate results. Fewer intermediate results indicate $\mathcal{P}$ will have a smaller chance to be full and flushed to DRAM, which can improve the overall performance.

Based on the above observation, Batch-DFS follows the idea of DFS. In DFS, the intermediate results are stored in a stack, and we always process its top element first. Similarly, we regard the buffer area $\mathcal{P}$ as a stack, and we always fetch a batch of paths from its top.

The details of Batch-DFS are shown in Algorithm \ref{alg:batch-dfs}. The inputs of the algorithm include the buffer area $\mathcal{P}$ and batch size threshold $\Theta$, and its output is a batch of paths $\mathcal{P'}$ that are going to be put into the processing area. We first initialize $cnt$ with 0 (Line 2), where $cnt$ denotes the number of intermediate paths that will be processed next. Hence we must ensure $cnt \leq \Theta$. Treating $\mathcal{P}$ as a stack, we use $i$ to trace $\mathcal{P}$ from its top (Line 3). We continuously fetch paths from $\mathcal{P}$ into $\mathcal{P'}$ until $cnt$ reaches the threshold $\Theta$ (Line 16-17). Due to the fact that there might exist a ``\textbf{super node}" whose degree is larger than $\Theta$, it is necessary to process its neighbors by batch, or it will blow the processing area. We achieve this by maintaining two pointers for each path's successors (or out-going neighbors), namely \textbf{start neighbor} and \textbf{end neighbor} pointer. The two pointers are initialized as pointing to the path's first neighbor. 
In Line 5, we declare two pointers $ptr_1, ptr_2$ and initialize them as the end neighbor pointer of $\mathcal{P}[i]$. Then we assign the last neighbor pointer of $\mathcal{P}[i]$ to $ptr_{last}$ (Line 6). If $ptr_1 + \Theta - cnt < ptr_{last}$ holds, then we set $ptr_2$ to position $ptr_1 + \Theta - cnt$, which indicates that the remaining space in the processing area is not enough to hold all successors of $\mathcal{P}[i]$. Thereby we can only load a batch of its successors (Line 7-8). If processing area has enough space, we just set $ptr_2$ to $ptr_{last}$ (Line 9-10). Then we put $\mathcal{P}[i]$ into $\mathcal{P'}$ and update the start and end neighbor pointer of $\mathcal{P}[i]$ with $ptr_1$ and $ptr_2$, respectively (Line 11-12). When we fetch $\mathcal{P}[i]$'s successors (Algorithm \ref{alg:pefp}, Line 5), we will perform this operation according to $ptr_1$ and $ptr_2$.

Overall, \pefp is a ``hybrid" algorithm that combines the merits of BFS and DFS. In other words, \pefp exerts the parallel ability of FPGA through BFS and saves on-chip memory through DFS-based batch processing. Consequently, \pefp successfully tames both computational hardness and FPGA on-chip memory bottleneck.

\begin{figure}
\vspace{-0.2cm}
\centering
\includegraphics[width=0.9\columnwidth]{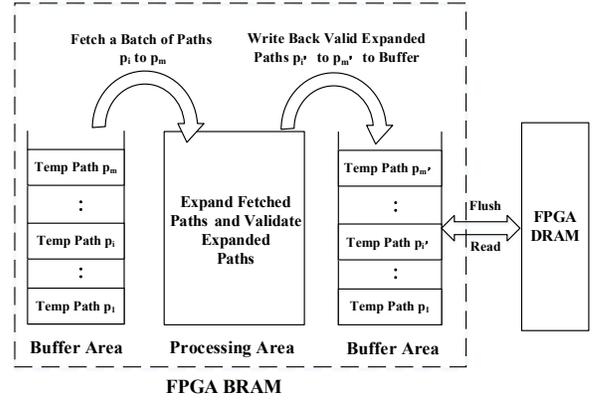}
\vspace{-0.2cm}
\caption{Overview of FPGA Batch Processing}
\label{fig:Overview-FPGA}
\end{figure}

\subsection{Basic Pipeline of Verification}
\label{subsec:basic-pipeline}
The read/write path operation can be easily pipelined on FPGA. The bottleneck remains in how to pipeline the verification of each expanded path. In this subsection, we will demonstrate a basic pipeline technique for path verification. Note that we call path verification and path validity check interchangeably in this paper.

\begin{figure}
\vspace{0.0cm}
\centering
\includegraphics[width=0.9\columnwidth]{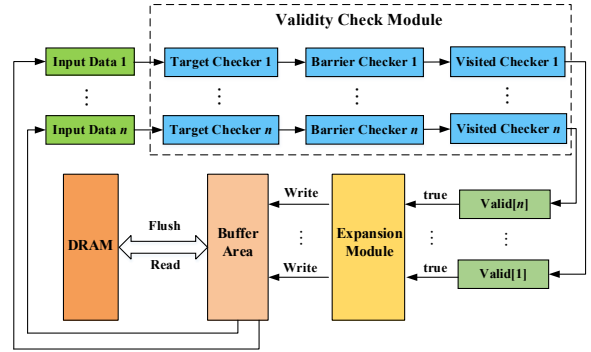}
\vspace{-0.1cm}
\caption{Basic Pipeline of Path Verification on FPGA}
\label{fig:Basic-FPGA}
\end{figure}

As illustrated in Fig.~\ref{fig:Basic-FPGA}, given an intermediate path $\mathcal{P'}[i]$, its one-hop successor $nbr$ and barrier $bar[nbr]$ (we call them input data $i$ in Fig.~\ref{fig:Basic-FPGA}), the basic verification module consists of three consecutively executed stages, namely (1) target check stage, (2) barrier check stage and (3) visited check stage. The input data of this module must include all the information required by the three stages according to Algorithm \ref{alg:verify}. The target check and barrier check can be finished in $O(1)$ time, while the visited check can be finished in $O(k)$ time without using hash set. We can unroll the visited check loop that has constant loop bound $k$, thus the time cost of visited check on FPGA can be reduced from $O(k)$ to $O(1)$. 

In this design, the verification for each input data is pipelined, while the three stages inside the module cannot be executed in parallel, because only when the input data passes the current stage can it move to the next stage.

\subsection{Optimized Pipeline of Verification by Data Separation}
\label{subsec:opt-pipeline}
% The acceleration power of FPGA mainly comes from its pipeline and dataflow optimizations. Pipeline optimization is usually applied to loops to make the loop body execute in parallel, where the premise is that there are no data dependencies between the current iteration and the next. Dataflow optimization is usually used in functional level to make the functions execute in parallel, where the data transfer between two functions is performed in streaming fashion. Therefore, we should carefully design the algorithm to ensure that each module can be fully pipelined inside, and different modules can be executed in parallel through dataflow, which is non-trivial.

For the basic design, although it can pipeline the input data verification, the inner stages of the verification module are executed one by one, which cannot exert the parallel ability of FPGA. We observe that the bottleneck is caused by data dependencies among the stages; that is, the current stage cannot be executed until the previous stage is finished. Motivated by this, we separate the input data according to the stages so that each stage can do its own computation independently. Then we merge the outputs of the three stages to get the final verification result for each expansion. 

\begin{figure}
\vspace{-0.2cm}
\centering
\includegraphics[width=0.9\columnwidth]{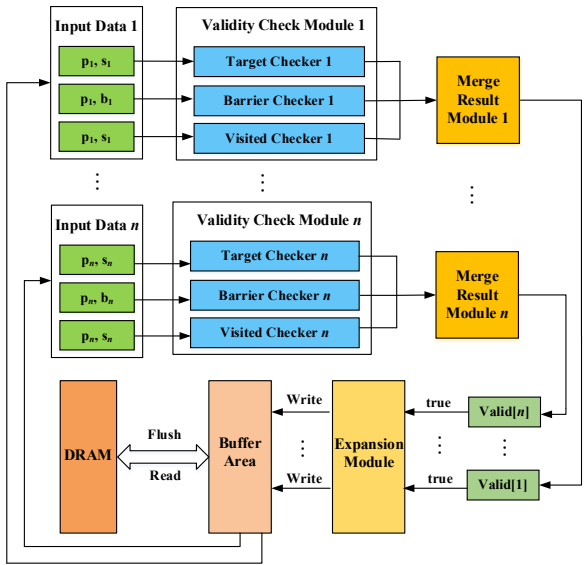}
\vspace{-0.2cm}
\caption{Optimized Pipeline of Path Verification with Data Separation on FPGA}
\label{fig:Opt-FPGA}
\end{figure}

More specifically, the input data of the path validity check module is separated into path $p_i$, successor $s_i$, and barrier $b_i$, as illustrated in Fig.~\ref{fig:Opt-FPGA}. In this design, we send $p_i$ and $s_i$ to the target check and visited check stage, and send $p_i$ and $b_i$ to the barrier check stage. Consequently, there are no data dependencies among the stages. We also apply dataflow optimization to the stages, suggesting that each stage can start its computation once it receives input data without waiting for the previous stage to finish. As a result, the three stages can be executed in parallel. All we need to do is merge the validity check results of the three stages, which improves the overall performance of \pefp.

\section{Experiment}
\label{sec:experiment}

\begin{table}
    \centering
    \begin{tabular}{c c c c c c c}
      \hline
      Dataset &Name   & $|V|$ & $|E|$ & $d_{avg}$ & $D$  & $D_{90}$
      \\ \hline
       Reactome   &RT & 6.3K   & 147K   & 46.64      & 24    & 5.39
        \\ 
        soc-Epinions1  &SE & 75K & 508K & 13.42 & 14 &5
        \\ 
        Slashdot0902  &SD & 82K & 948K & 23.08 & 12  & 4.7
        \\
        Amazon &AM & 334K & 925K & 6.76 & 44 & 15
        \\ 
        twitter-social &TS & 465K & 834K & 3.86 & 8 & 4.96
        \\
        Baidu  &BD & 425K & 3M & 15.8 & 32 & 8.54
        \\
        BerkStan &BS & 685K & 7M & 22.18 & 208 & 9.79
        \\ 
        web-google &WG & 875K & 5M & 11.6 & 24 & 7.95
        \\
        Skitter &SK & 1.6M & 11M & 13.08 & 31 & 5.85
        \\
        WikiTalk &WT & 2M & 5M & 4.2 & 9 & 4
        \\ 
        LiveJournal &LJ & 4M & 68M & 28.4 & 16 & 6.5
        \\
        DBpedia &DP & 18M & 172M & 18.85  & 12 & 4.98
        \\ \hline
    \end{tabular}
%\vspace{-2mm}
\caption{Statistics of Datasets}
\label{tb:dataset}
\vspace{-2mm}
\end{table}

\begin{figure*}[htb]
\vspace{-0.2cm}
	\newskip\subfigtoppskip \subfigtopskip = -0.1cm
	\newskip\subfigcapskip \subfigcapskip = -0.1cm
	\begin{minipage}[b]{\linewidth}
		\centering
		\includegraphics[width=0.25\linewidth]{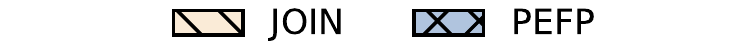}%
	\end{minipage}
	
     \centering
    \subfigure[Amazon]{
     \includegraphics[width=0.23\linewidth]{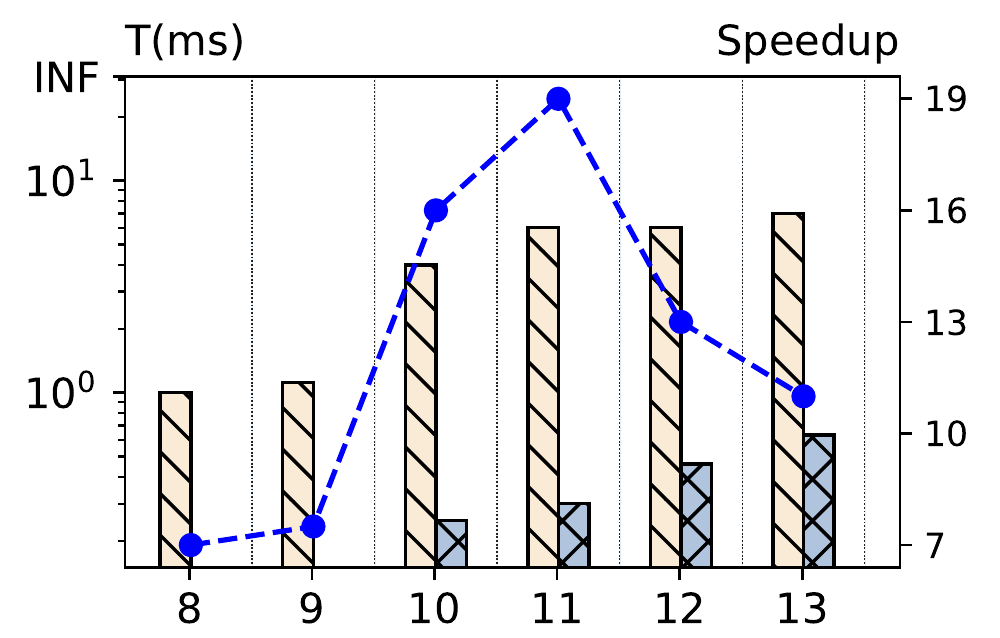}
	 \label{fig:amazon:query_time}
     }
     \subfigure[WikiTalk]{
     \includegraphics[width=0.23\linewidth]{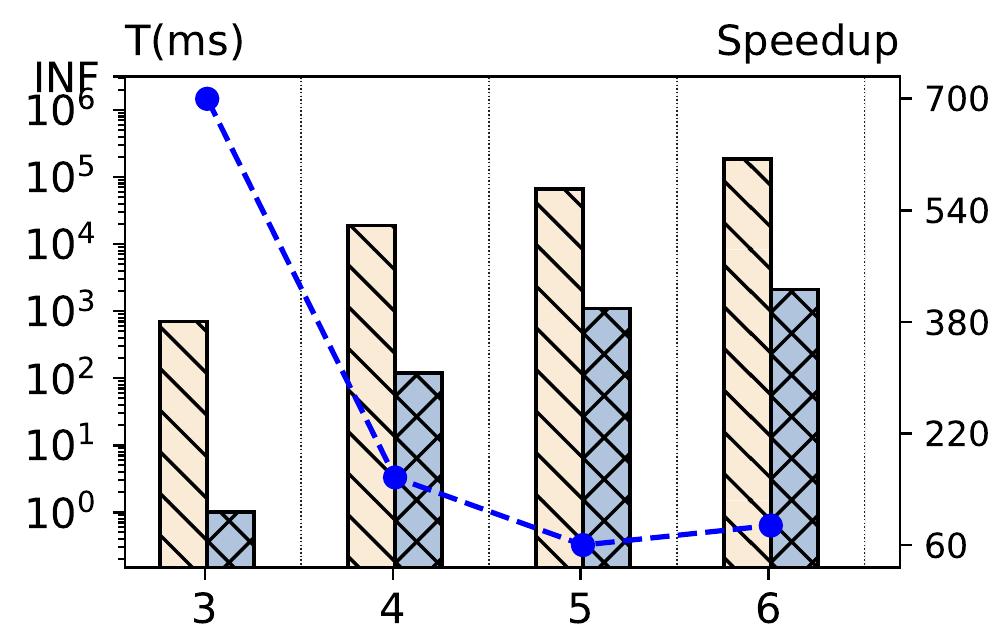}%
     \label{fig:wiki-talk:query_time}
     }
     \subfigure[Skitter]{
     \includegraphics[width=0.23\linewidth]{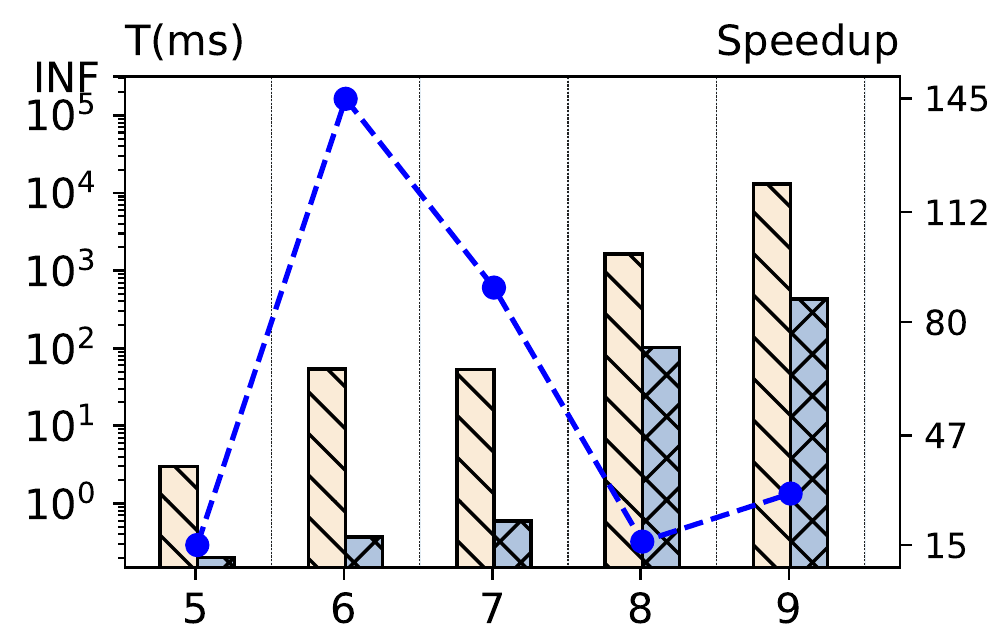}%
	 \label{fig:skitter:query_time}
	 }
	 \subfigure[twitter-social]{
     \includegraphics[width=0.23\linewidth]{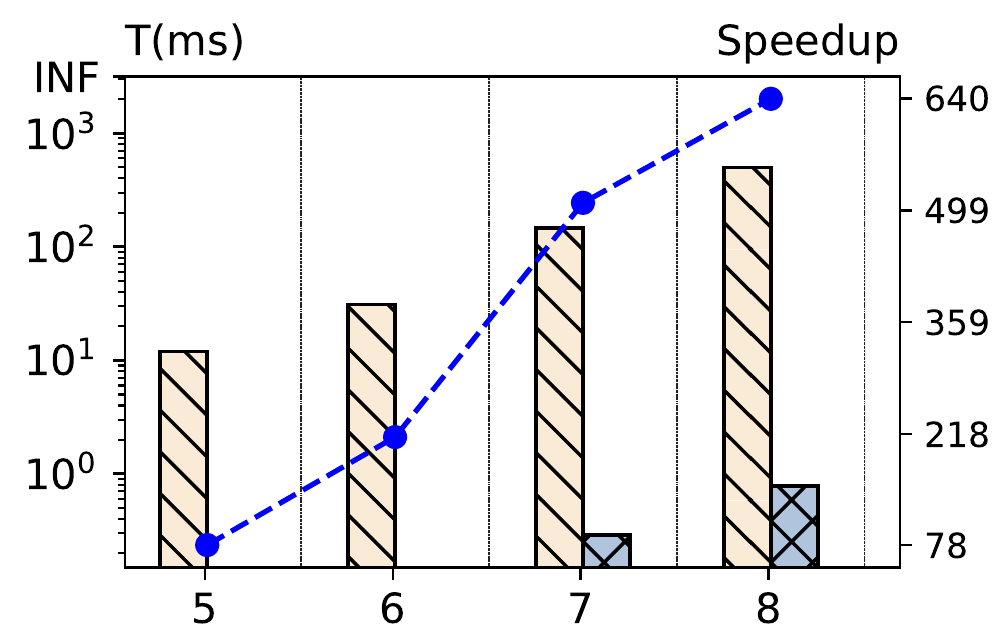}%
     \label{fig:twitter-social:query_time} 
     }
     
      \subfigure[Baidu]{
     \includegraphics[width=0.23\linewidth]{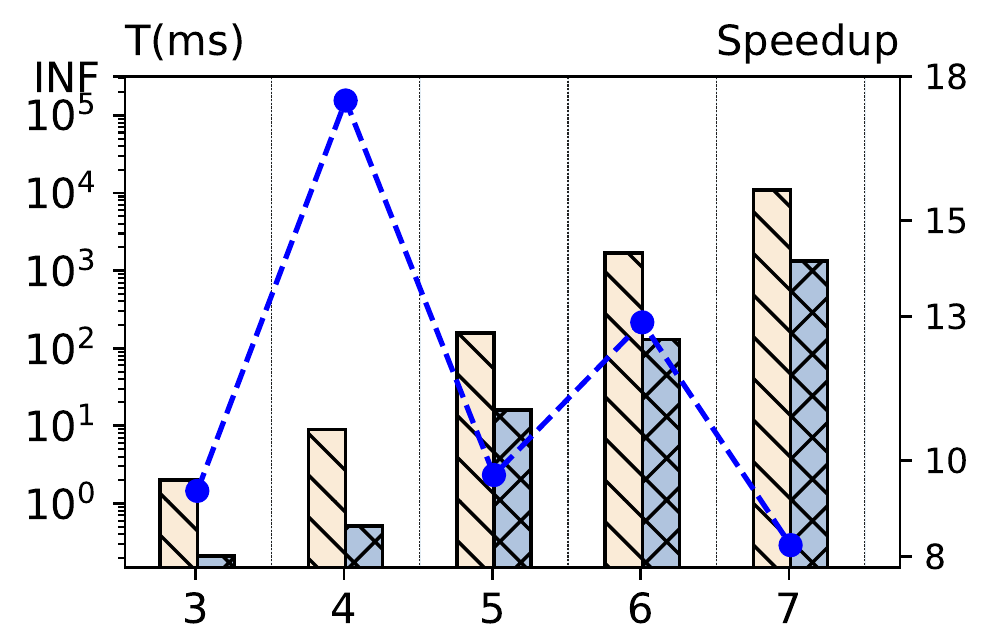}
	 \label{fig:baidu:query_time}
     }
     \subfigure[BerkStan]{
     \includegraphics[width=0.23\linewidth]{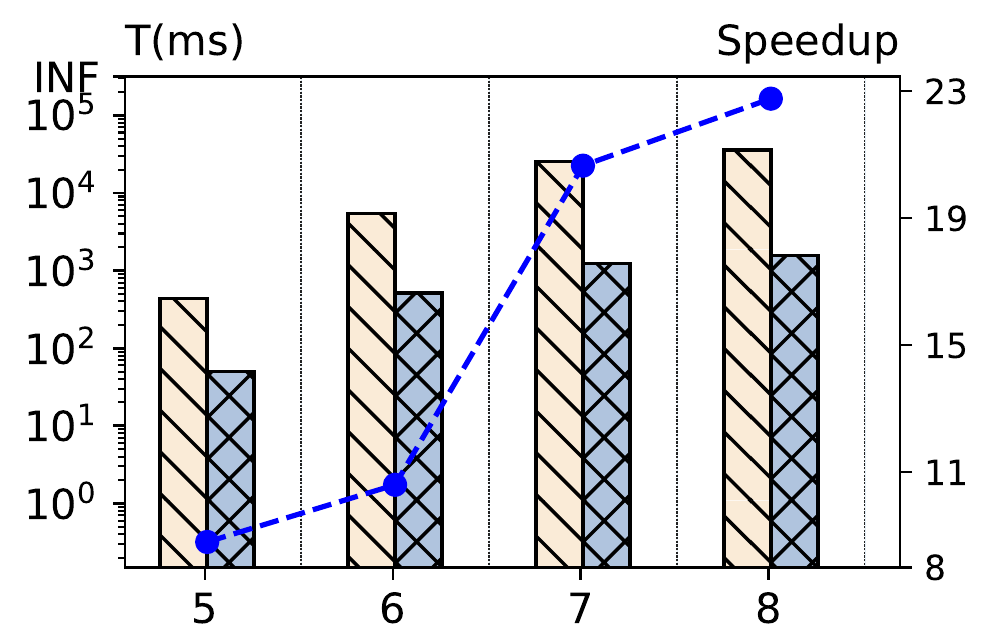}%
     \label{fig:berkstan:query_time}
     }
     \subfigure[soc-Epinions1]{
     \includegraphics[width=0.23\linewidth]{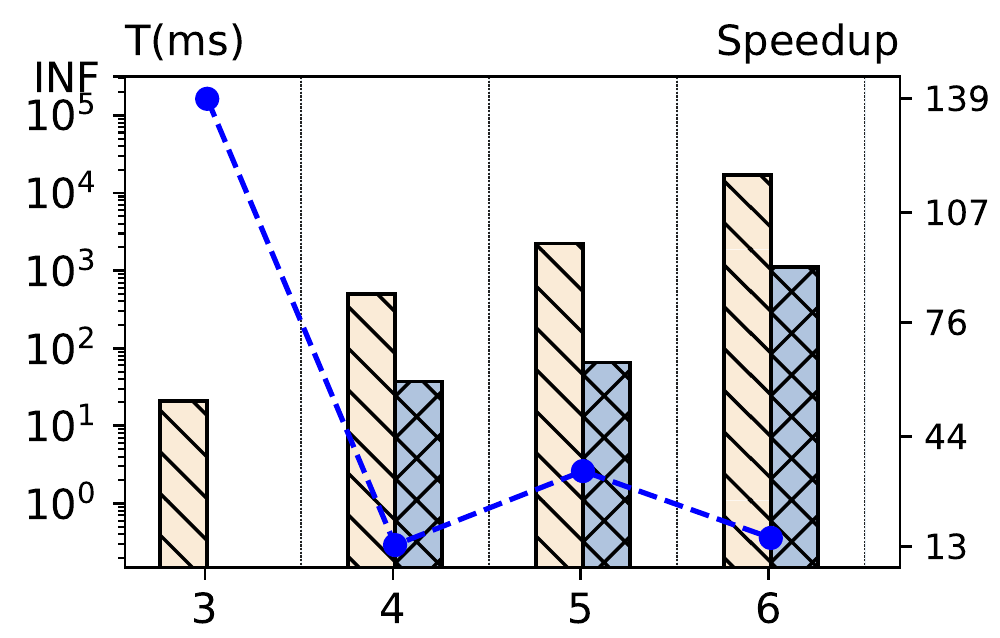}%
	 \label{fig:epinions:query_time}
	 }
	 \subfigure[web-google]{
     \includegraphics[width=0.23\linewidth]{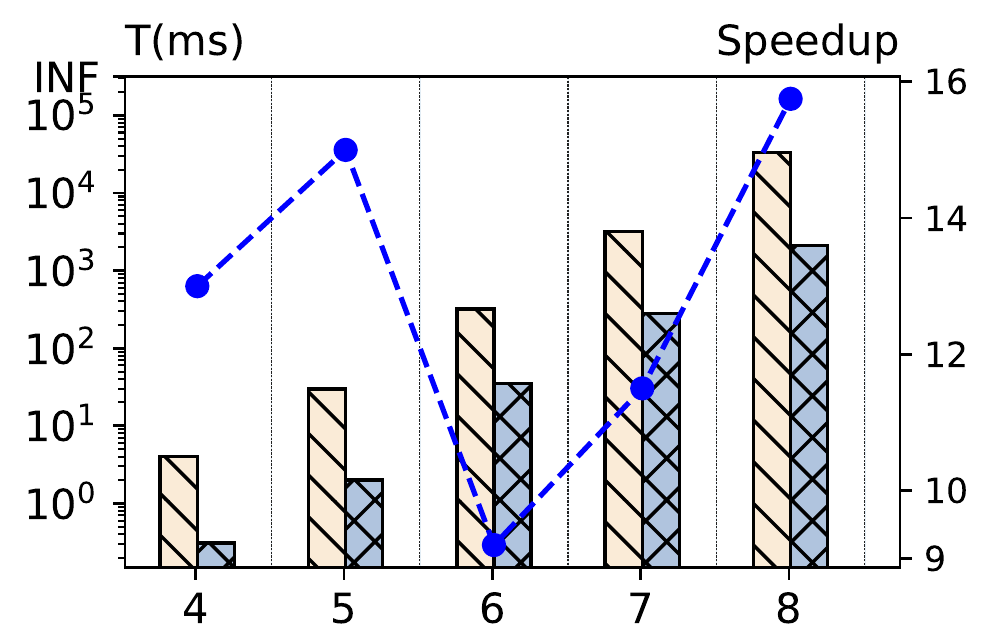}%
     \label{fig:web-google:query_time} 
     }
     
      \subfigure[LiverJournal]{
     \includegraphics[width=0.23\linewidth]{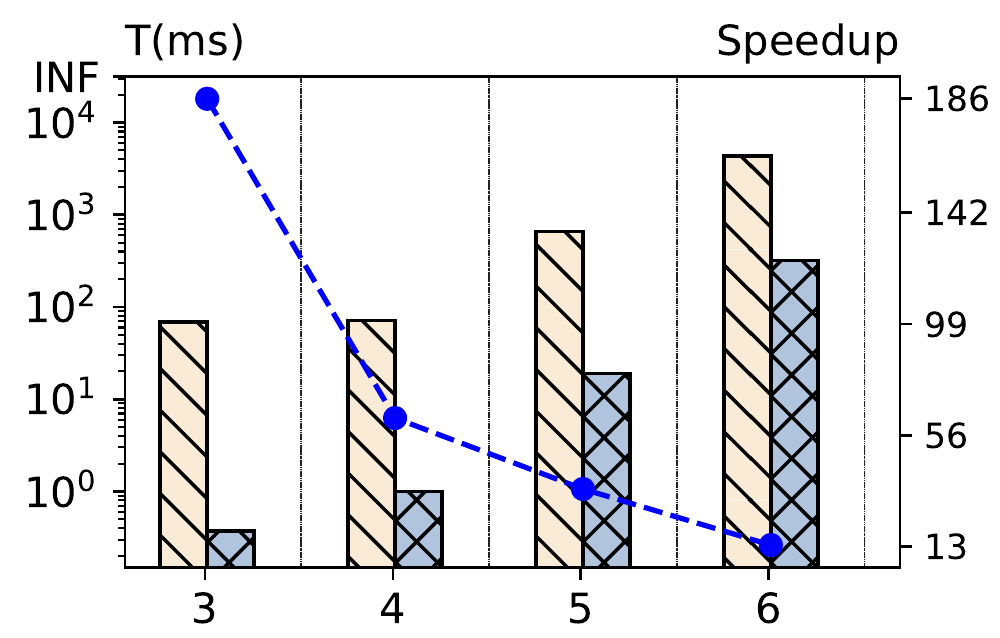}
	 \label{fig:lj:query_time}
     }
     \subfigure[Reactome]{
     \includegraphics[width=0.23\linewidth]{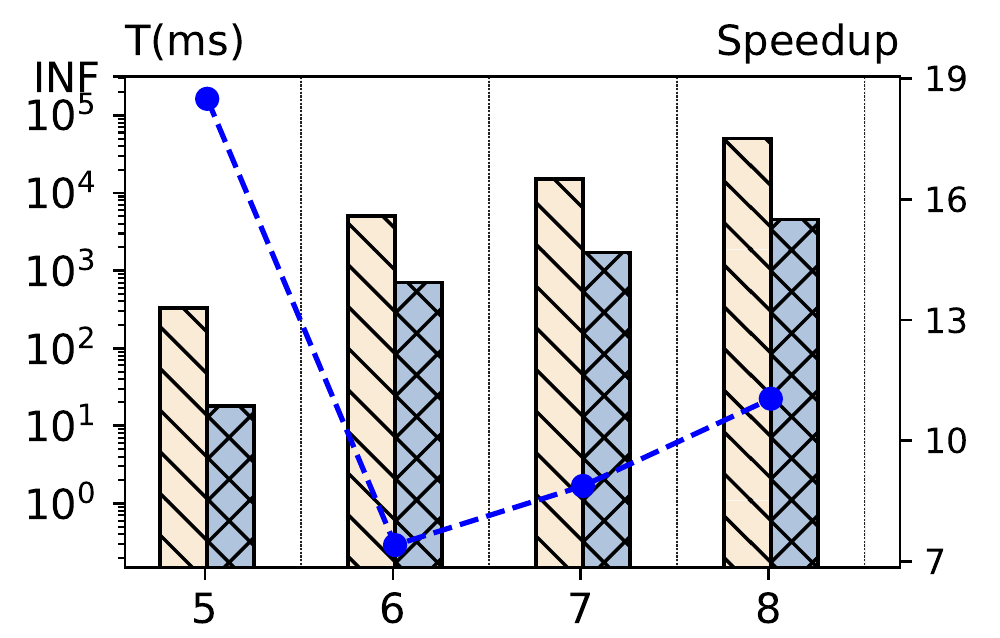}%
     \label{fig:reactome:query_time}
     }
     \subfigure[Slashdot0902]{
     \includegraphics[width=0.23\linewidth]{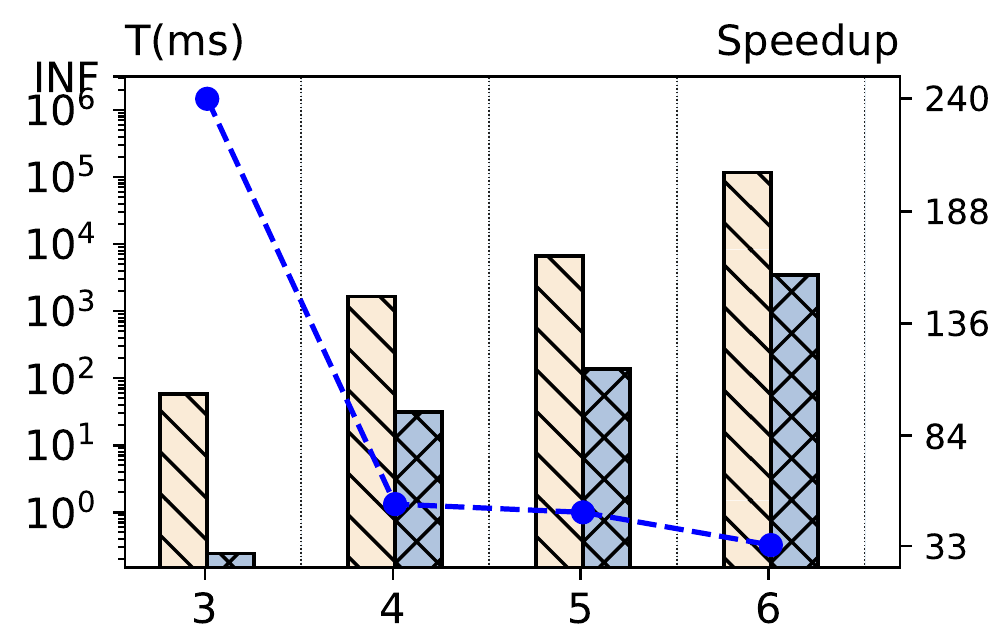}%
     \label{fig:slash:query_time}
     }
     \subfigure[DBpedia]{
     \includegraphics[width=0.23\linewidth]{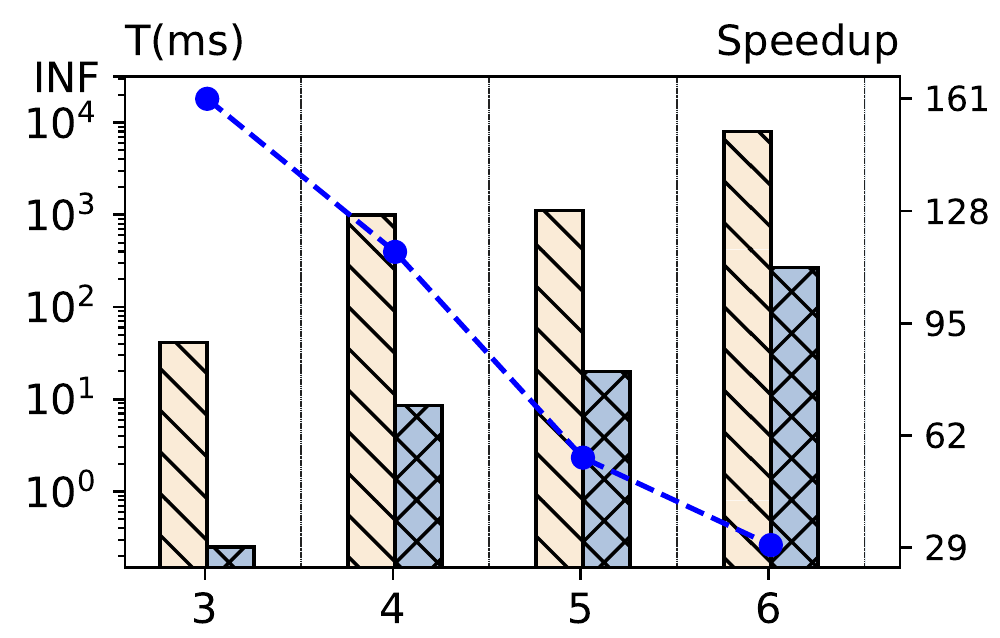}%
     \label{fig:dbpedia:query_time}
     }
\vspace{-2mm}
\caption{Query Processing Time of Tuning $k$ for All Datasets}
\label{fig:tuning_k_for_qt}
\end{figure*}

\begin{figure*}[htb]
\vspace{-0.2cm}
	\newskip\subfigtoppskip \subfigtopskip = -0.1cm
	\newskip\subfigcapskip \subfigcapskip = -0.1cm
	
	\newskip\subfigtoppskip \subfigtopskip = -0.1cm
	\newskip\subfigcapskip \subfigcapskip = -0.1cm
	\begin{minipage}[b]{\linewidth}
		\centering
		\includegraphics[width=0.25\linewidth]{legend.pdf}%
	\end{minipage}
	
     \centering
    \subfigure[Amazon]{
     \includegraphics[width=0.23\linewidth]{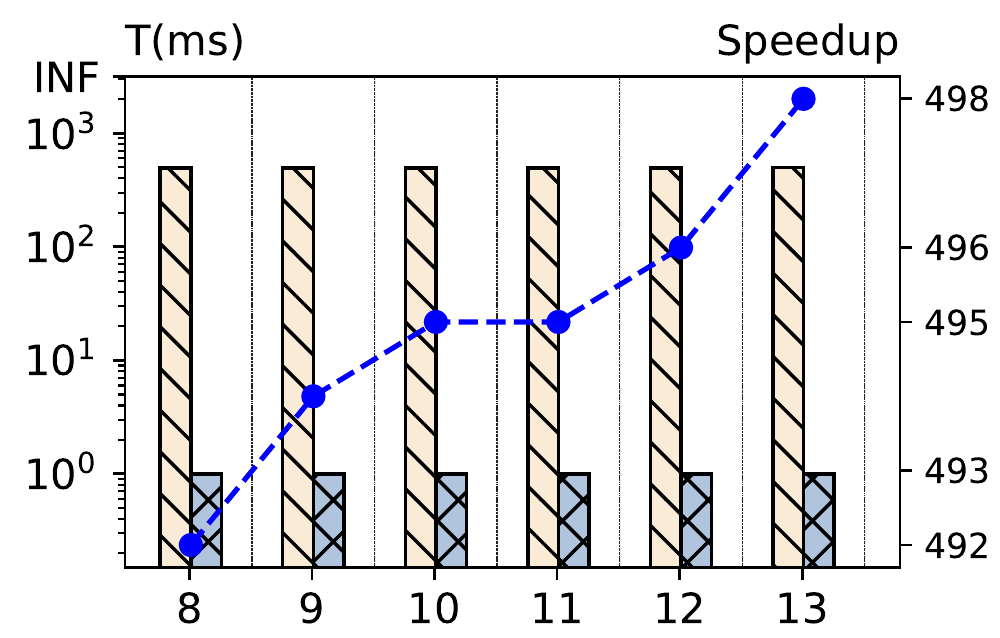}
	 \label{fig:amazon:pre_time}
     }
     \subfigure[WikiTalk]{
     \includegraphics[width=0.23\linewidth]{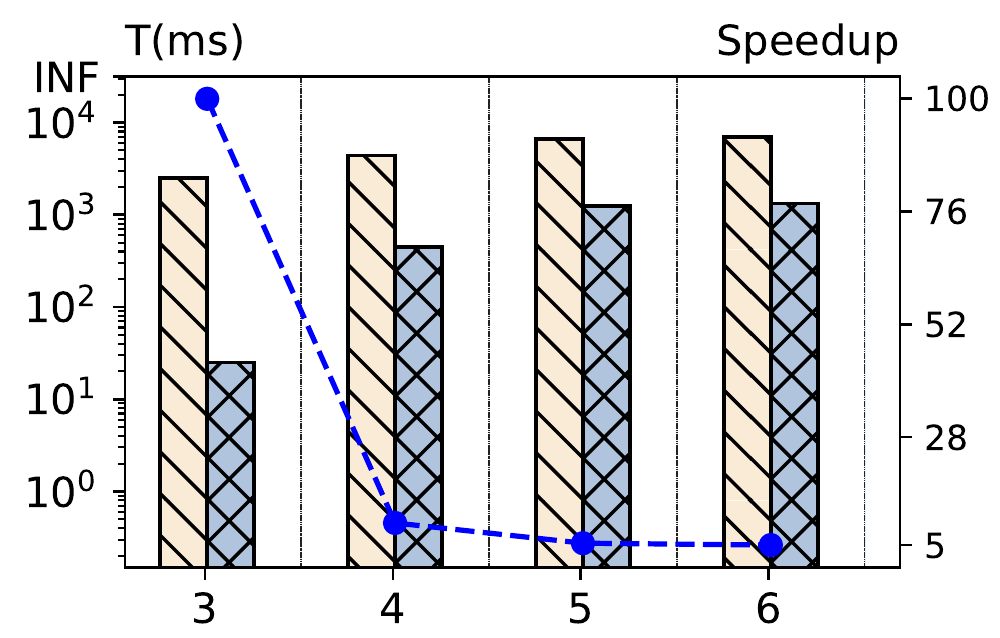}%
     \label{fig:wiki-talk:pre_time}
     }
     \subfigure[Skitter]{
     \includegraphics[width=0.23\linewidth]{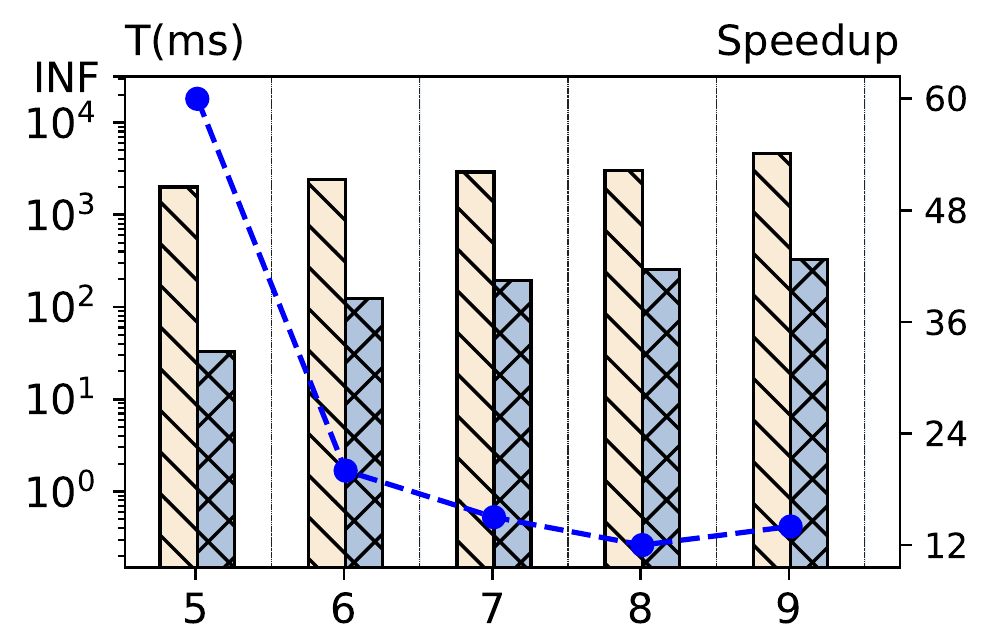}%
	 \label{fig:skitter:pre_time}
	 }
	 \subfigure[twitter-social]{
     \includegraphics[width=0.23\linewidth]{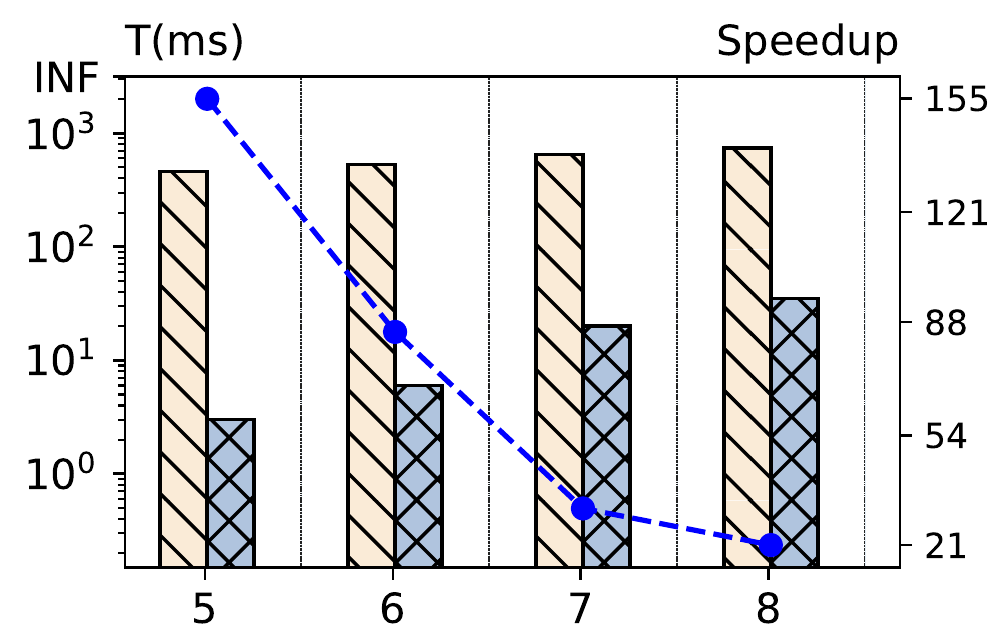}%
     \label{fig:twitter-social:pre_time} 
     }
\vspace{-2mm}
\caption{Preprocessing Time of Tuning $k$ for Different Datasets}
\label{fig:tuning_k_for_pre}
\end{figure*}

\begin{figure*}[htb]
\vspace{-0.2cm}
	\newskip\subfigtoppskip \subfigtopskip = -0.1cm
	\newskip\subfigcapskip \subfigcapskip = -0.1cm
	
	\newskip\subfigtoppskip \subfigtopskip = -0.1cm
	\newskip\subfigcapskip \subfigcapskip = -0.1cm
	\begin{minipage}[b]{\linewidth}
		\centering
		\includegraphics[width=0.25\linewidth]{legend.pdf}%
	\end{minipage}
	
     \centering
    \subfigure[Amazon]{
     \includegraphics[width=0.23\linewidth]{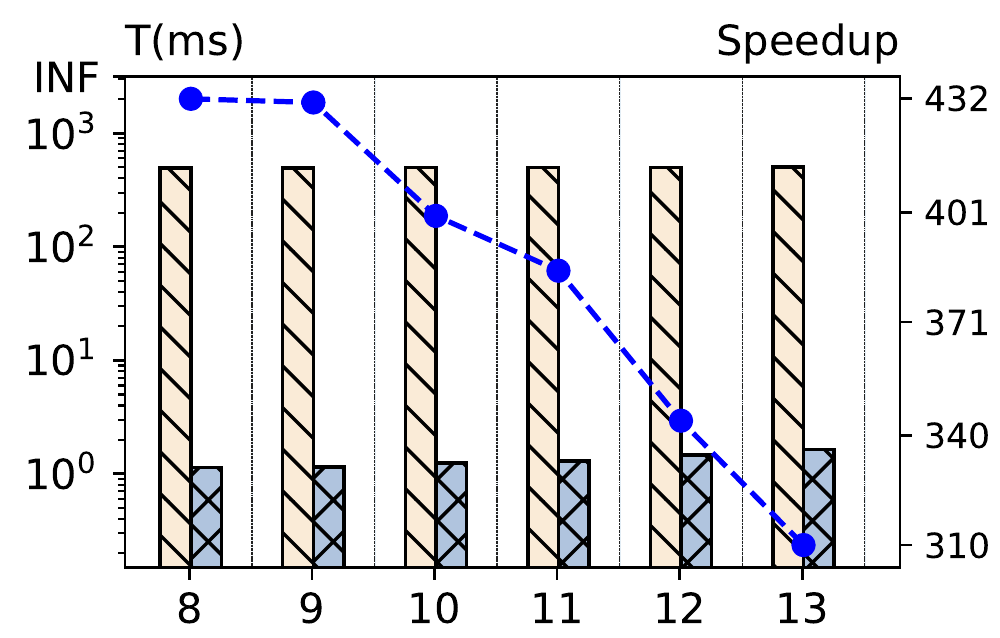}
	 \label{fig:amazon:total_time}
     }
     \subfigure[WikiTalk]{
     \includegraphics[width=0.23\linewidth]{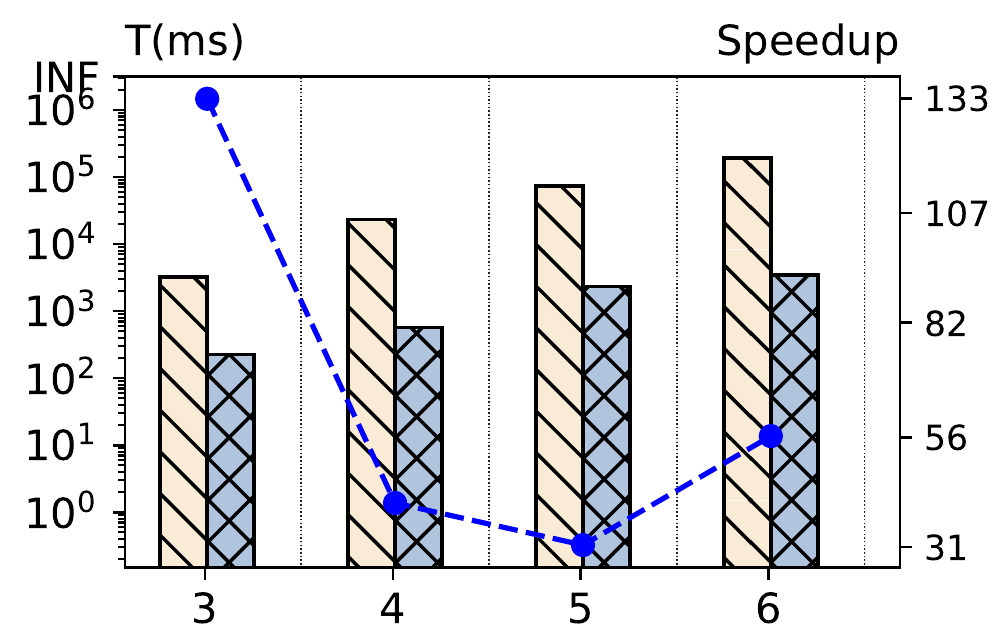}%
     \label{fig:wiki-talk:total_time}
     }
     \subfigure[Skitter]{
     \includegraphics[width=0.23\linewidth]{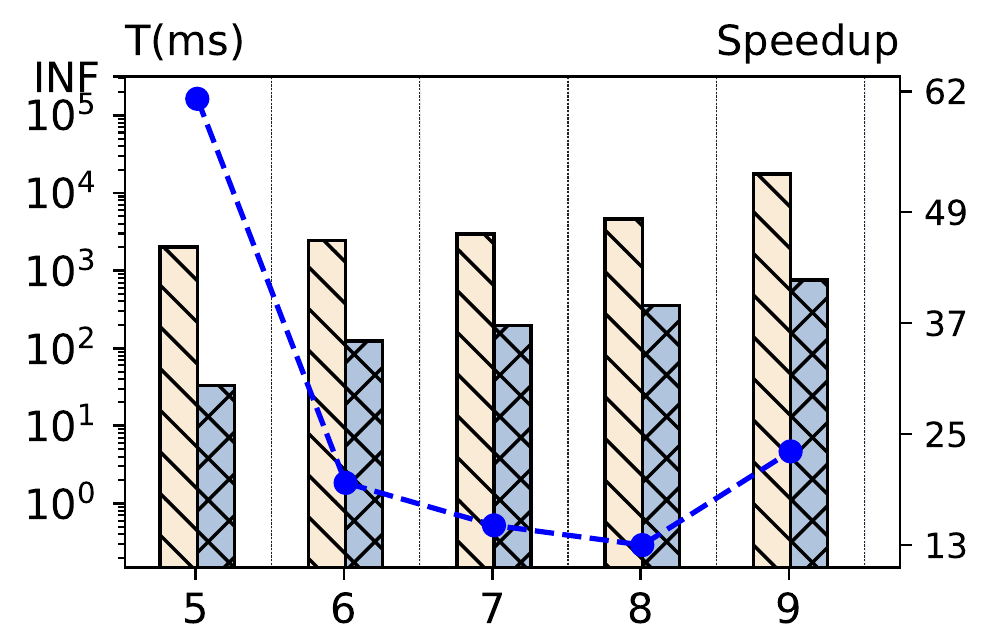}%
	 \label{fig:skitter:total_time}
	 }
	 \subfigure[twitter-social]{
     \includegraphics[width=0.23\linewidth]{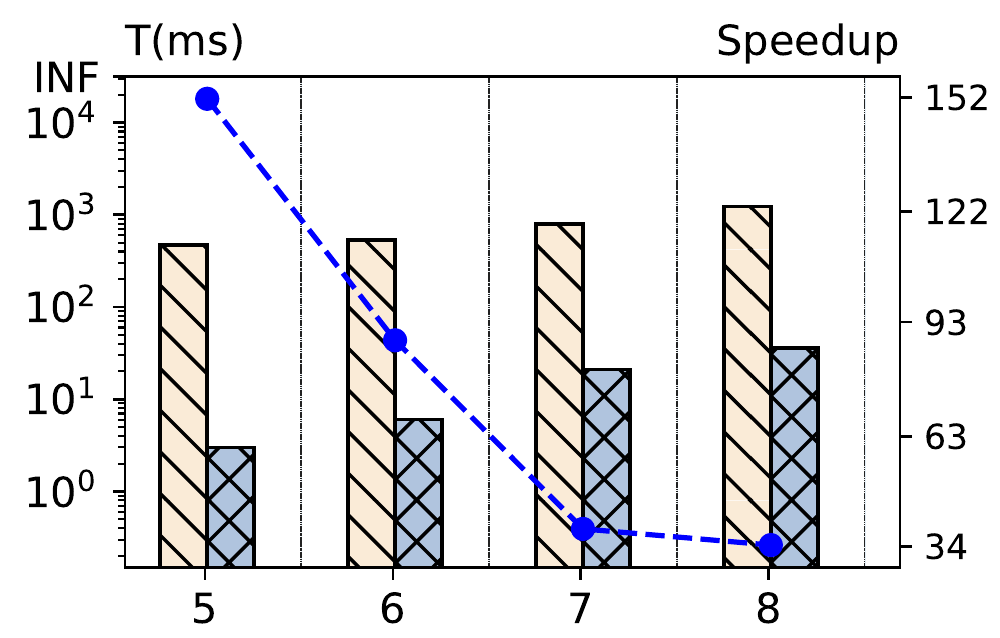}%
     \label{fig:twitter-social:total_time} 
     }
\vspace{-2mm}
\caption{Total Time of Tuning $k$ for Different Datasets}
\label{fig:tuning_k_for_total_time}
\end{figure*}

In this section, we conduct extensive experiments to evaluate the effectiveness and efficiency of our proposed algorithm \pefp. As discussed in Section~\ref{sec:related}, none of the related FPGA works can be directly adapted to solve $s$-$t$ $k$-path enumeration problem. Moreover, JOIN is the state-of-the-art algorithm based on DFS paradigm, and it is non-trivial to implement a parallel version of JOIN either on multi-core CPUs or on FPGA. Therefore, in this paper, we address the original JOIN algorithm as the baseline solution, and we compare the performance of \pefp with JOIN on a wide range of datasets. 

\subsection{Experiment Setup}
\vspace{1mm}
\noindent \textbf{Settings.} All experiments are conducted on a Ubuntu 16.04 machine, with 250GB memory, 10TB disk and 2 Intel Xeon CPUs (E5-2620 v4 2.10GHz, 8 cores). The proposed FPGA-based algorithm \pefp is implemented on Xilinx Alveo U200 card~\footnote{https://www.xilinx.com/products/boards-and-kits/alveo/u200.html} using Xilinx SDAccel~\footnote{https://www.xilinx.com/products/design-tools/software-zone/sdaccel.html}, where the FPGA card is equipped with 4 $\times$ 16GB off-chip DRAMs and runs at 300MHz. The code of JOIN is obtained from the authors in~\cite{qin2019towards}, which is implemented in standard C++ and compiled with g++ 5.5.0~\footnote{JOIN's source code:  https://github.com/zhengminlai/JOIN}.

\vspace{1mm}
\noindent \textbf{Datasets.} 
All datasets are downloaded from two public websites:
Konect~\footnote{http://konect.uni-koblenz.de/networks/} and SNAP~\footnote{http://snap.stanford.edu/data/}.
TABLE~\ref{tb:dataset} demonstrates detailed data descriptions. Note that $d_{avg}$ denotes the average degree, $D$ denotes the diameter and $D_{90}$ denotes the 90-percentile effective diameter of the graph.

\vspace{1mm}
\noindent \textbf{Metrics.} 
We randomly generate 1,000 query pairs \{$s$, $t$\} for each dataset with hop constraint $k$, where the source vertex $s$ could reach target vertex $t$ in $k$ hops. We then evaluate the average running time of the 1,000 queries, where each query's running time is obtained from an average of three runs. Note that for better presentation, we carefully set the range of $k$ for each dataset based on its topology. For instance, $k$ is set from 8 for Amazon, which is a rather sparse graph. Moreover, for each dataset, we have evaluated the time it takes to transfer the 1,000 queries and their corresponding data graphs (after preprocessing) from the host to FPGA DRAM at once, which is around 100ms$\sim$300ms. Hence, the average transfer time for each query is around 0.1ms$\sim$0.3ms, which can be ignored as the preprocessing time in host and query processing time on FPGA would dominate.
We denote query preprocessing time as $T_1$, query processing time as $T_2$, and total time as $T = T_1 + T_2$.
We evaluate $T_1, T_2$, and $T$ for both \pefp and JOIN, where the preprocessing of JOIN is introduced in Section~\ref{sec:soft_preprocessing}.

\subsection{Evaluate Query Processing Time}
\label{subsec:exp:query-processing}
\vspace{1mm}

In this experiment, we evaluate the query processing time of \pefp and JOIN on all 12 datasets by varying the hop constraint $k$, which is illustrated in Fig.~\ref{fig:tuning_k_for_qt}. The blue dotted line in that figure represents the speedup of \pefp over JOIN, which is same as the remaining experiments. We set the query time of an algorithm to \textit{INF} if it cannot finish in 10,000 seconds. 

\noindent \textbf{(1) Effect of $k$.} The results shown in Fig.~\ref{fig:tuning_k_for_qt} indicate that \pefp's query processing time outperforms JOIN on all datasets for fixed $k$. It is not surprising that the time grows exponentially w.r.t $k$ as $s$-$t$ $k$-path number grows exponentially w.r.t $k$~\cite{qin2019towards}. However, this does not hold for Amazon (Fig.~\ref{fig:amazon:query_time}) -- the time only grows marginally w.r.t $k$. An explanation would be that Amazon is an extremely small and sparse graph, and hence the number of reported results would be too small to see significant changes of query time when tuning $k$. Note that in BerkStan (Fig.~\ref{fig:berkstan:query_time}), the query time of $k=7$ is almost the same as the one of $k=8$. This makes sense as we find their reported number of paths are on the same order of magnitude, which is $10^6$. The same explanation can be applied to WikiTalk (Fig.~\ref{fig:wiki-talk:query_time}) with $k=5$ and $k=6$.

From the perspective of the acceleration ratio (or speedup) w.r.t $k$, there are some interesting findings from inspecting Fig.~\ref{fig:tuning_k_for_qt}. For most graphs like Amazon, Baidu, BerkStan and Reactome, the acceleration ratio remains rather stable, which is around 10$\times$ to $20\times$ speedup. Note that the greatest difference between JOIN and \pefp is that JOIN is a DFS-based algorithm with carefully designed pruning technique BC-DFS, while \pefp is a BFS-based parallel algorithm on FPGA with a less delicate pruning technique (e.g., barrier check shown in \ref{sec:hardwareImpl}). When $k$ is small, query processing is dominated by the expansion rather than the verification procedure, where the expansion can be fully pipelined in \pefp. Therefore, a substantial speedup can be observed when $k$ is small in most graphs (e.g., $>600\times$ speedup in WikiTalk with $k=3$).

Another intriguing fact, as illustrated in Fig.~\ref{fig:twitter-social:query_time}, is the speedup of twitter-social tends to increase when $k$ ranges from 5 to 8. This is because twitter-social is a graph with a very low diameter, which is 4.96 for 90$\%$ of the graph as demonstrated in TABLE~\ref{tb:dataset}. Nevertheless, the minimal $k$ set in twitter-social is 5, thus the pruning power of both JOIN's BC-DFS and \pefp's barrier check is almost zero. Under such condition, the query time is dominated by expansion; hence \pefp shows its considerable superiority over JOIN. We can explain the dramatic upsurge of speedup in Skitter from $k=5$ to $k=6$ in similar way; that is, as the $D_{90}$ of Skitter is 5.85, the pruning power of BC-DFS becomes rather weak when $k$ changes from 5 to 6.

\vspace{1mm}
\noindent \textbf{(2) Effect of Dataset.} 
It is apparent that for a given $k$, the query processing time varies in datasets with different graph topologies. What stands out in Fig.~\ref{fig:tuning_k_for_qt} is that twitter-social's query time is much more than Amazon's (e.g., $k=8$). Although their numbers of vertices and edges are similar according to TABLE~\ref{tb:dataset}, the diameter of Amazon is 44 while the diameter of twitter-social is only 8. This implies that twitter-social is a graph with considerable local density. Consequently, for a given $k$, the query time of twitter-social is much more than that of Amazon. However, the speedup in Amazon is significantly less than that in twitter-social. The reason is that the number of intermediate results in Amazon is too low to observe a substantial acceleration ratio. The same explanation can be applied to Skitter with $k=5$.

Baidu is a smaller dataset with similar average degree compared to Skitter. Nevertheless, for a given $k$ (e.g., $k=6$), the query time of Baidu is much more than Skitter's. This is because there exist some extremely dense subgraphs in Baidu. In addition, the acceleration ratio of Baidu is less competitive than Skitter's, suggesting that \pefp tends to have a greater speedup in sparse graphs than in dense graphs. A possible reason is that the pruning power of JOIN is stronger in dense graphs, which tames the acceleration ratio brought by the parallelism of \pefp.

Overall, benefited from the huge parallelism offered by FPGA and the reduced search space by induced subgraph, \pefp outperforms JOIN by more than 1 order of magnitude by average, and up to 2 orders of magnitude in query processing time.

\begin{figure*}[htb]
\vspace{-0.2cm}
	\newskip\subfigtoppskip \subfigtopskip = -0.1cm
	\newskip\subfigcapskip \subfigcapskip = -0.1cm
	\begin{minipage}[b]{\linewidth}
		\centering
		\includegraphics[width=0.25\linewidth]{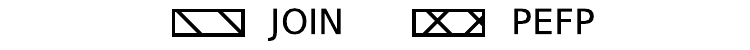}%
	\end{minipage}
	
     \centering
    \subfigure[Dataset Group 1]{
     \includegraphics[width=0.27\linewidth]{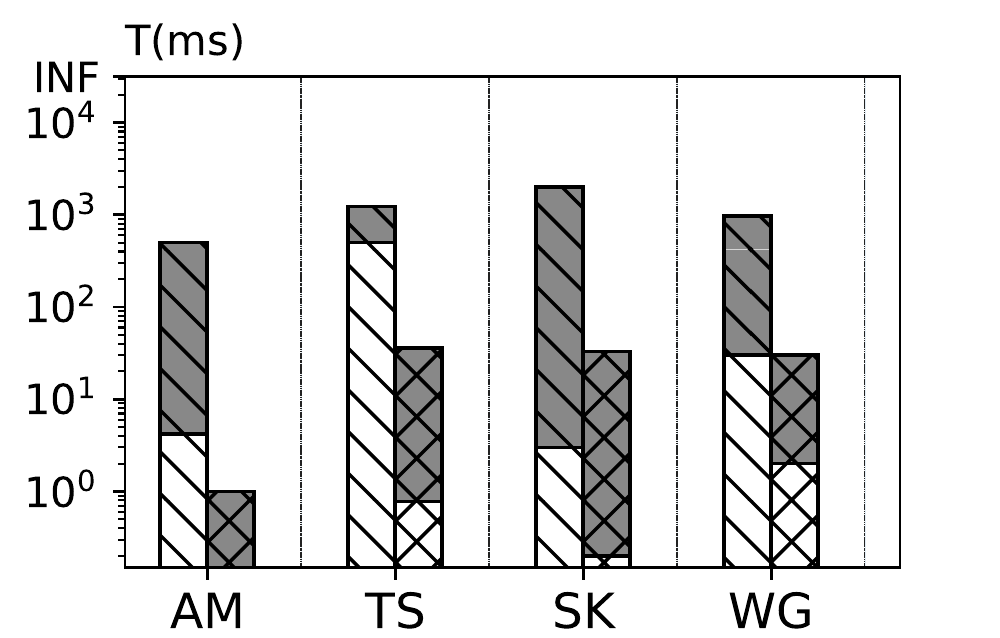}
	 \label{fig:avg_total_time:group1}
     }
     \subfigure[Dataset Group 2]{
     \includegraphics[width=0.27\linewidth]{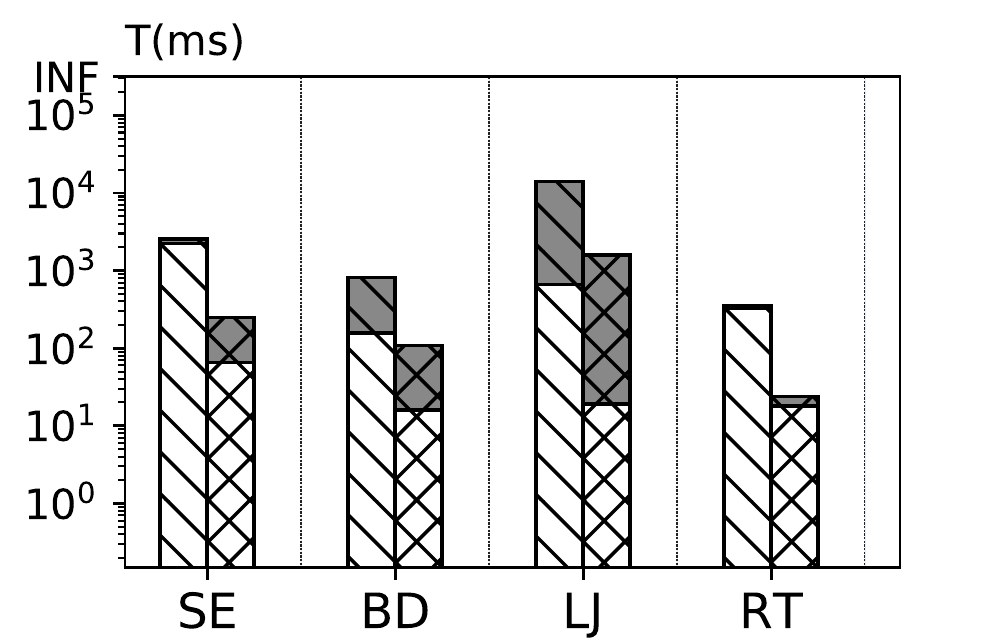}%
     \label{fig:avg_total_time:group2}
     }
     \subfigure[Dataset Group 3]{
     \includegraphics[width=0.27\linewidth]{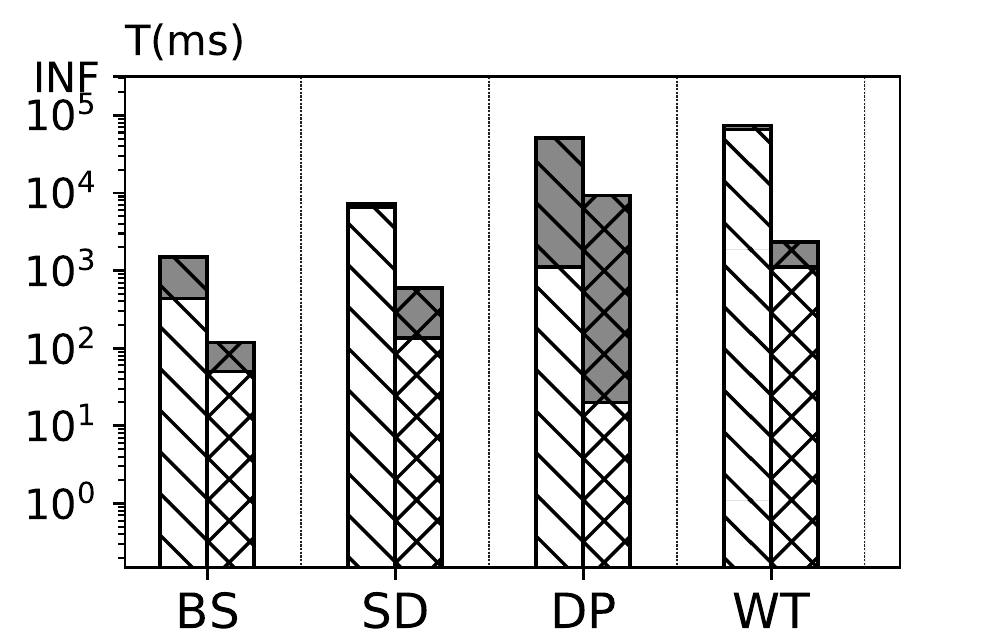}%
	 \label{fig:avg_total_time:group3}
	 }
\vspace{-2mm}
\caption{Average Total Time of All Datasets with $k=5$}
\label{fig:Avg_total_time}
\end{figure*}

\subsection{Evaluate Query Preprocessing Time}
\label{subsec:exp:preprocessing}

In Fig.~\ref{fig:tuning_k_for_pre}, we evaluate the query preprocessing time of \pefp and JOIN by varying the hop constraint $k$ on four datasets with different topologies, namely Amazon, WikiTalk, Skitter and twitter-social.

Fig.~\ref{fig:tuning_k_for_pre} shows that \pefp outperforms JOIN in all datasets w.r.t $k$. Particularly, in Fig.~\ref{fig:amazon:pre_time}, \pefp is 495$\times$ faster than JOIN by average, and the acceleration ratio does not change dramatically as $k$ increases. The main reason is that Amazon is a very small and sparse graph, and minor tuning of $k$ does not affect the preprocessing time of both JOIN and \pefp in such graphs. In addition, it is reported in Fig.~\ref{fig:amazon:pre_time} that JOIN's complicated preprocessing procedures are rather costly in small and sparse graphs compared with \pefp. 

As shown in Fig.~\ref{fig:wiki-talk:pre_time}, Fig.~\ref{fig:skitter:pre_time}, and
Fig.~\ref{fig:twitter-social:pre_time}, we can expect more than 10$\times$ acceleration ratio on average. Specifically, \pefp can achieve more than $100\times$ speedup when $k$ is small(e.g., $k=5$ in Fig.~\ref{fig:twitter-social:pre_time}), since JOIN's preprocessing cost is much more expensive than \pefp's in this case. Furthermore, the acceleration ratio of the three datasets tends to decrease w.r.t $k$, for JOIN's preprocessing time w.r.t $k$ is more stable than \pefp in these datasets. Nevertheless, JOIN fails to outperform \pefp as its preprocessing needs to perform $k$-hop BFS and expensive set intersections in computing middle vertices cut~\cite{qin2019towards}, while \pefp only needs to perform $(k-1)$-hop BFS as introduced in Section~\ref{sec:soft_preprocessing}. As a result, \pefp outperforms JOIN by more than 1 order of magnitude by average, and up to 2 orders of magnitude in preprocessing time.

\subsection{Evaluate Total Time}
\label{subsec:exp:total-time}
In this subsection, we evaluate the total running time of a given query, where the total time is the sum of query preprocessing and query processing time. As shown in Fig.~\ref{fig:tuning_k_for_total_time}, we investigate the total time of \pefp and JOIN by varying the hop constraint $k$ on four datasets, namely Amazon, WikiTalk, Skitter and twitter-social. Then we report the total time of all datasets with $k=5$ in Fig.~\ref{fig:Avg_total_time}, where the white part represents query processing time, and the grey part denotes preprocessing time.

\vspace{1mm}
\noindent \textbf{(1) Effect of $k$.} In Fig.~\ref{fig:tuning_k_for_total_time}, we report the total time of JOIN and \pefp on four datasets by varying $k$. The acceleration ratio of Amazon tends to decrease when $k$ ranges from $10$ to $13$. This makes sense as \pefp's total time w.r.t $k$ grows faster than JOIN's in an extremely sparse graph. For the other three datasets, a similar trend of speedup is observed. Particularly, the speedup is substantial with a small $k$. Nevertheless, when $k$ increases, the speedup tends to first decrease, and then remain stable. This is because when $k$ is small, the total time of JOIN is dominated by preprocessing time, and hence we can expect a considerable speedup as discussed in Section~\ref{subsec:exp:preprocessing}. However, this significant advantage brought by preprocessing will be paid off by query processing when increasing $k$. 

\vspace{1mm}
\noindent \textbf{(2) Effect of Dataset.} The total time of all datasets is illustrated in Fig.~\ref{fig:Avg_total_time}. We set $k=8$ for \textit{Amazon} and \textit{twitter-social} to achieve similar performance with other graphs while $k=5$ for the remaining graphs. What stands out in Fig.~\ref{fig:Avg_total_time} is that both JOIN and \pefp's total time are dominated by preprocessing time in sparse graphs like Amazon and Skitter, while the total time of JOIN in twitter-social is dominated by query processing time because of the dataset's local density topology. It is worth mentioning that the graph density is a key factor influencing total time -- when the graph is sparse, the total time is dominated by preprocessing time, and vice versa.

In short, the results in this experiment show that \pefp outperforms JOIN by more than 1 order of magnitude by average, and up to 2 orders of magnitude in total time.

\begin{figure}
	\newskip\subfigtoppskip \subfigtopskip = -0.02cm
	\newskip\subfigcapskip \subfigcapskip = -0.1cm
	\begin{minipage}[b]{\columnwidth}
		\centering
		\includegraphics[width=0.5\columnwidth]{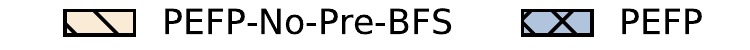}%
	\end{minipage}
	
	\subfigure[BerkStan]{
    	\label{fig:pre-bfs:berkstan} 
    	\includegraphics[width=0.46\columnwidth]{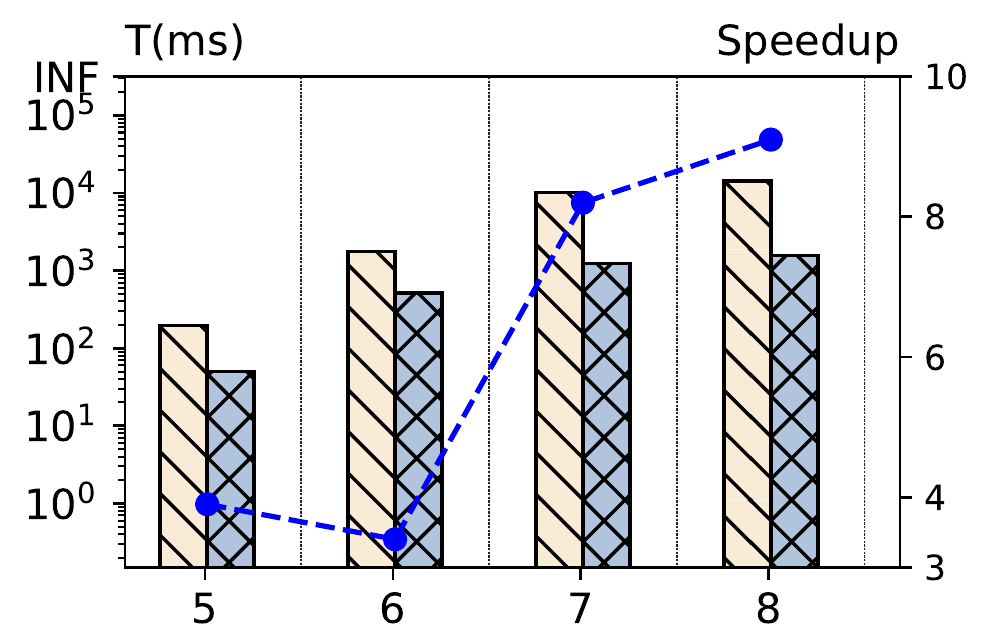}
    }
  	\subfigure[Baidu]{
    	\label{fig:pre-bfs:baidu}
    	\includegraphics[width=0.46\columnwidth]{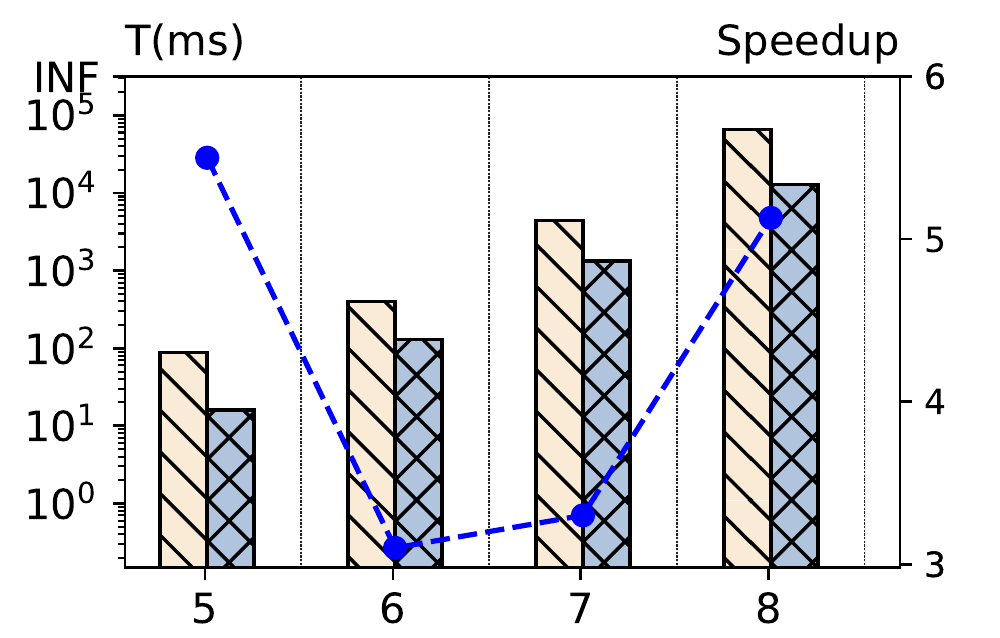}
    }
\vspace{-0.2cm}
\caption{Evaluation of Pre-BFS Technique Tuning $k$}
\label{fig:pre-bfs}
\end{figure}

\subsection{Evaluate Efficiency of Pre-BFS}
\label{subsec:exp:pre-bfs}
\vspace{1mm}
As demonstrated in Fig.~\ref{fig:pre-bfs}, we evaluate the efficiency of our proposed preprocessing algorithm Pre-BFS on BerkStan and Baidu, where \textbf{PEFP-No-Pre-BFS} denotes the \pefp algorithm without Pre-BFS. It is shown that Pre-BFS can achieve $3\times$ to $9\times$ speedup in the two datasets. As we mentioned in Section~\ref{sec:soft_preprocessing} and Section~\ref{sec:hardwareImpl}, Pre-BFS can improve the performance of \pefp in two ways. First, it significantly reduces the search space by removing invalid nodes that will not be contained in any $s$-$t$ $k$-path. Second, the subgraph it extracts is much smaller than the original graph, making it possible for FPGA to cache the whole subgraph on BRAM; thus, it is necessary to apply Pre-BFS optimization for $s$-$t$ $k$-path enumeration.

\begin{table}
    \centering
    \begin{tabular}{c c c c c c c}
      \hline
      Dataset &$l=2$ & $l=3$ & $l=4$ & $l=5$ & $l=6$ & $l=7$
      \\ \hline
        Baidu &3117	&17346	&10033	&4522	&1064	&0
        \\
        BerkStan &9374	&14376	&10678	&7991	&5114	&0
        \\ 
        WikiTalk &52498	&103544	&63935	&13207	&1198	&0
        \\ 
        LiveJournal  &276802 &351396 &299003 &165018 &11027	&0
        \\ \hline
    \end{tabular}
%\vspace{-2mm}
\caption{Number of Newly Generated Intermediate Paths When Doing One-hop Expansion with 1,000 paths for Different Path Length $l$ with $k=8$}
\label{tb:num-of-intermediate-paths}
\vspace{-2mm}
\end{table}

\begin{figure}
	\newskip\subfigtoppskip \subfigtopskip = -0.02cm
	\newskip\subfigcapskip \subfigcapskip = -0.1cm
	\begin{minipage}[b]{\columnwidth}
		\centering
		\includegraphics[width=0.5\columnwidth]{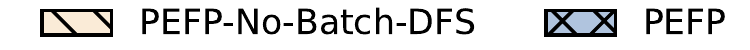}%
	\end{minipage}
	
	\subfigure[BerkStan]{
    	\label{fig:dfs-batch:berkstan} 
    	\includegraphics[width=0.46\columnwidth]{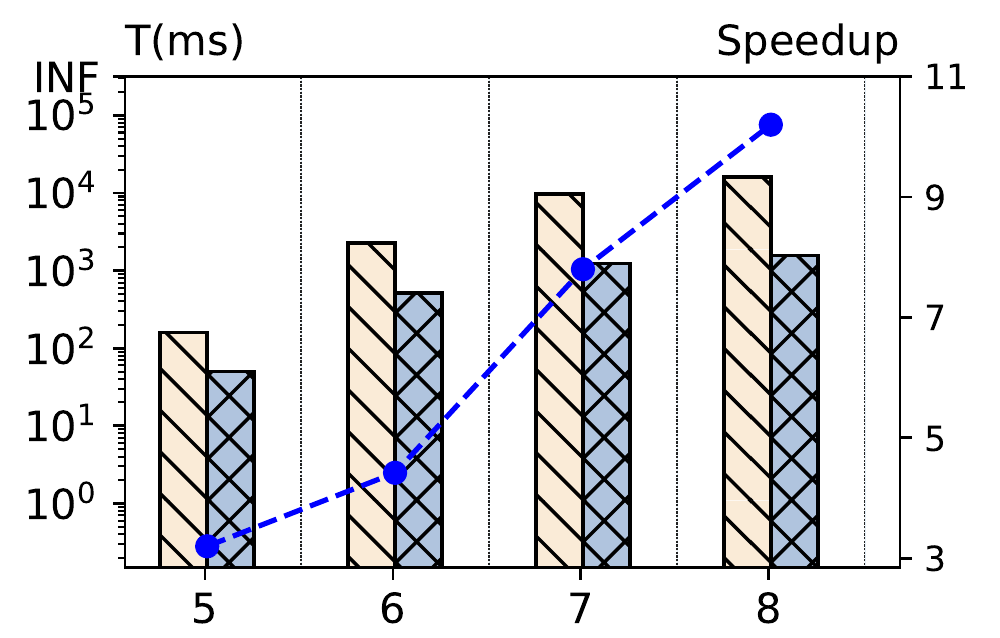}
    }
  	\subfigure[Baidu]{
    	\label{fig:dfs-batch:baidu}
    	\includegraphics[width=0.46\columnwidth]{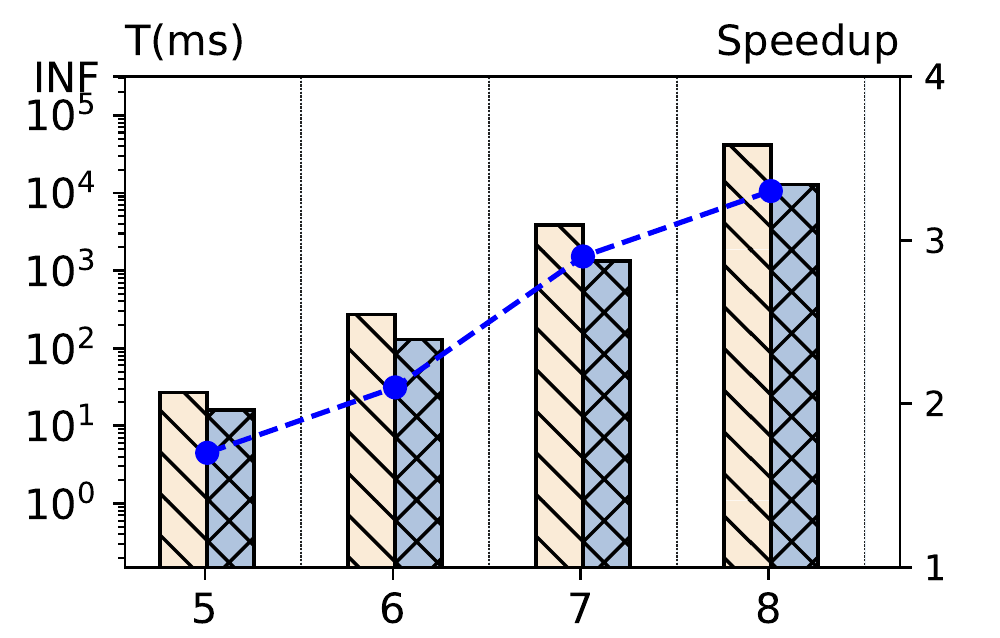}
    }
\vspace{-0.2cm}
\caption{Evaluation of Batch-DFS Technique Tuning $k$}
\label{fig:dfs-batch}
\end{figure}

\subsection{Evaluate Efficiency of Batch-DFS}
\label{subsec:exp:dfs-batch}
\vspace{1mm}
\noindent \textbf{(1) Number of Intermediate Paths.} To better illustrate the intuition of Batch-DFS, for each path length $l \in [2, k-1]$ (we set $k$ to 8 in this experiment), we randomly pick 1,000 paths to do one-hop expansion and evaluate the number of newly generated intermediate paths on four datasets -- Baidu, BerkStan, WikiTalk and LiveJournal. The experimental results are presented in TABLE~\ref{tb:num-of-intermediate-paths}, which shows that given $k=8$, for two small path lengths $l_1 = 2$, $l_2 = 3$, the number of newly produced paths tends to increase. We attribute this to the fact that, when $l$ is small, the pruning power of path length is rather weak, while the chance of touching the high degree nodes tends to increase w.r.t $l$. Nevertheless, the pruning power of hop constraint is getting stronger when $l$ becomes large -- the number of newly generated paths tends to decrease when $l > 3$. Specifically, it will generate 0 intermediate paths when $l = k - 1 = 7$. Therefore, the experimental results demonstrate the effectiveness of Batch-DFS which follows the art of \textit{``always process a batch of the longest paths first"} to save memory.

\noindent \textbf{(2) Query Time.} The efficiency evaluation of our proposed Batch-DFS technique on BerkStan and Baidu is shown in Fig.~\ref{fig:dfs-batch}, where \textbf{PEFP-No-Batch-DFS} denotes the \pefp algorithm without Batch-DFS. Instead, we use First-In-First-Out (FIFO) batching order to replace Batch-DFS, which is \textit{``always process a batch of the shortest paths first"}. The results in that figure show that Batch-DFS can achieve $2\times$ to $10\times$ speedup. Moreover, we can see that the speedup for BerkStan is higher than Baidu's. This is reasonable as the number of intermediate results of BerkStan is larger than Baidu's, which brings more I/O cost without Batch-DFS.

\begin{figure}
	\newskip\subfigtoppskip \subfigtopskip = -0.02cm
	\newskip\subfigcapskip \subfigcapskip = -0.1cm
	\begin{minipage}[b]{\columnwidth}
		\centering
		\includegraphics[width=0.5\columnwidth]{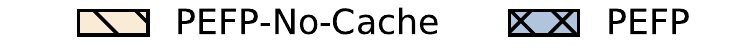}%
	\end{minipage}
	
	\subfigure[Reactome]{
    	\label{fig:cache:reactome} 
    	\includegraphics[width=0.46\columnwidth]{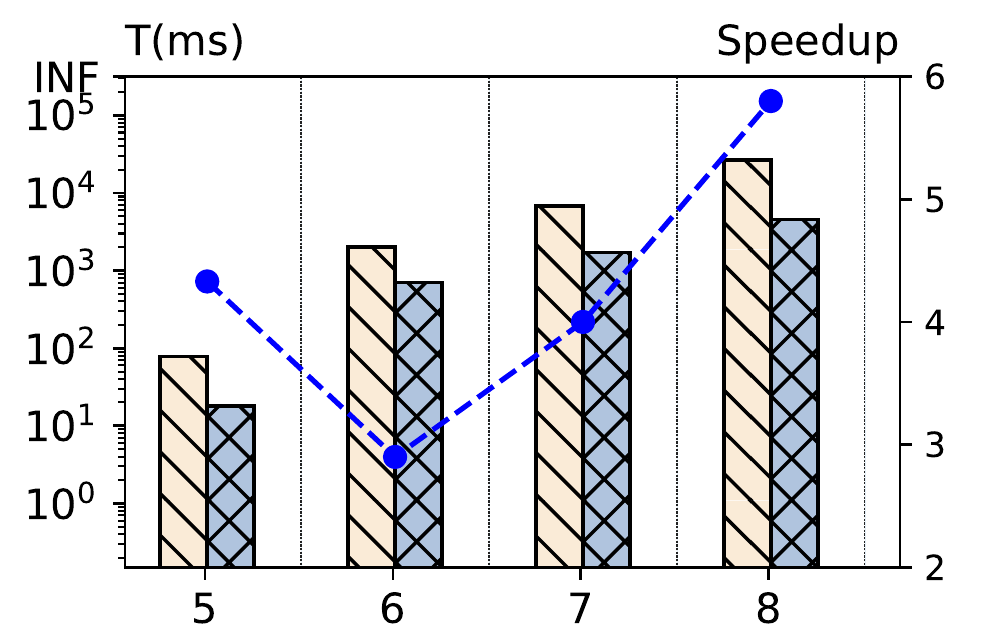}
    }
  	\subfigure[web-google]{
    	\label{fig:cache:google}
    	\includegraphics[width=0.46\columnwidth]{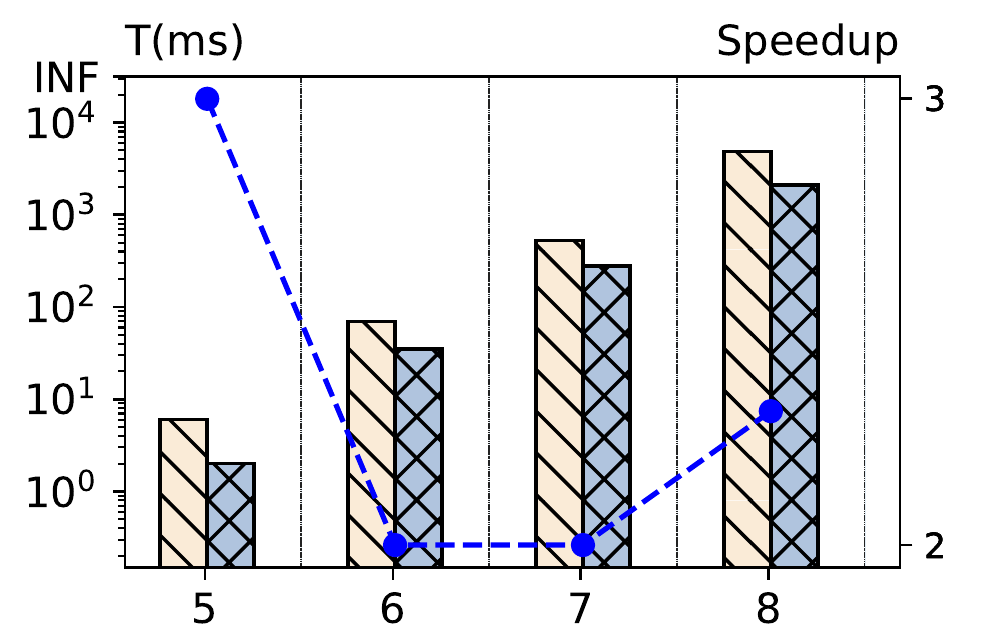}
    }
\vspace{-0.2cm}
\caption{Evaluation of Caching Technique Tuning $k$}
\label{fig:cache}
\end{figure}

\subsection{Evaluate Efficiency of Caching}
\label{subsec:exp:caching}
\vspace{1mm}
In Fig.~\ref{fig:cache}, we evaluate the efficiency of our proposed caching techniques on Reactome and web-google, where we use \textbf{PEFP-No-Cache} to denote the \pefp algorithm without caching techniques. It is shown that caching can achieve more than $2\times$ speedup by average, and up to $6\times$ speedup for PEFP-No-Cache. It is worth mentioning that caching results in better speedup in Reactome than in web-google. This makes sense as Reactome is a much denser graph than web-google, which incurs more vertex and edge data accesses to DRAM; hence its performance is significantly affected. 

\begin{figure}
	\newskip\subfigtoppskip \subfigtopskip = -0.02cm
	\newskip\subfigcapskip \subfigcapskip = -0.1cm
	\begin{minipage}[b]{\columnwidth}
		\centering
		\includegraphics[width=0.5\columnwidth]{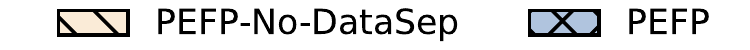}%
	\end{minipage}
	
	\subfigure[Reactome]{
    	\label{fig:dataflow:reactome} 
    	\includegraphics[width=0.46\columnwidth]{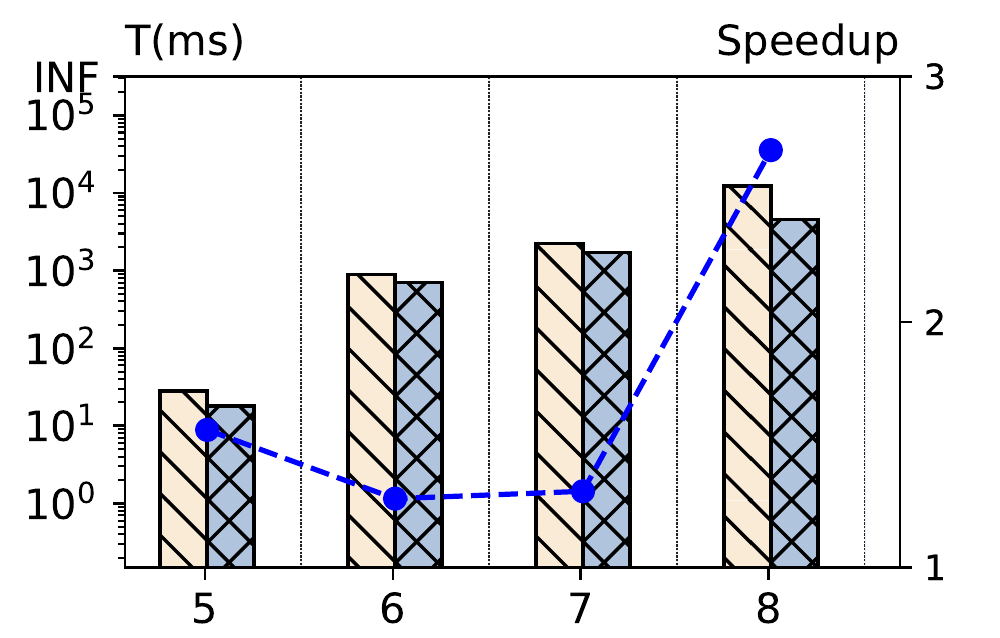}
    }
  	\subfigure[web-google]{
    	\label{fig:dataflow:google}
    	\includegraphics[width=0.46\columnwidth]{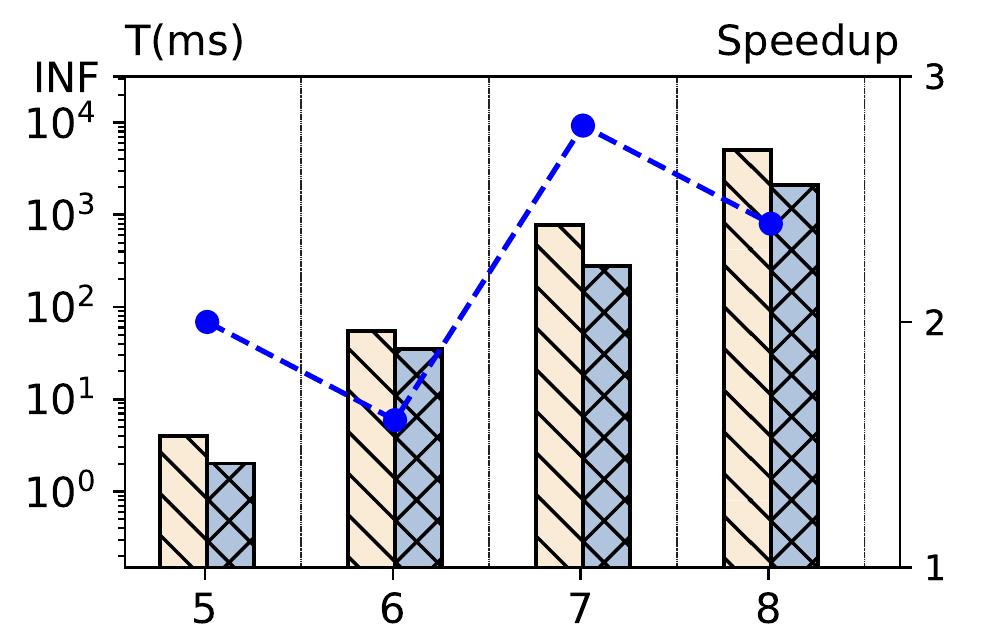}
    }
\vspace{-0.2cm}
\caption{Evaluation of Data Separation Technique Tuning $k$}
\label{fig:datasep}
\end{figure}

\subsection{Evaluate Efficiency of Data Separation}
\label{subsec:exp:datasep}
\vspace{1mm}
As illustrated in Fig.~\ref{fig:datasep}, we evaluate the efficiency of our proposed data separation technique on Reactome and web-google, where \textbf{PEFP-No-DataSep} denotes the \pefp algorithm without data separation technique. The results in that figure show that data separation can achieve up to $3\times$ speedup. This is because data separation enables dataflow optimization for the path verification module such that its inner stages can be executed in parallel, which improves the overall performance.
\section{Conclusion}
\label{sec:conclusion}

In this paper, we propose the first FPGA-based algorithm \pefp to efficiently solve the $s$-$t$ $k$-path enumeration problem. On the host side, we develop the preprocessing algorithm Pre-BFS to reduce the search space. On the FPGA side, we first propose a novel DFS-based batching and caching technique to improve the system latency by reducing read/write operations from/to FPGA DRAM. Then, a data separation technique for the path verification module is developed, which enables its inner stages to be executed in parallel. We conduct extensive experiments on 12 real-world datasets, whose results show that \pefp outperforms the state-of-the-art algorithm JOIN by more than 1 order of magnitude by average, and up to 2 orders of magnitude in terms of preprocessing time, query processing time and total time, respectively.
\section*{Acknowledgment}
Zhengmin Lai is supported by  The National Key R\&D Program of China under grant 2018YFB1003504. Xuemin Lin is supported by The National Key R\&D Program of China under grant 2018YFB1003504, NSFC61232006, ARC DP200101338, ARC DP180103096 and ARC DP170101628. Wenjie Zhang is supported by ARC DP200101116 and DP180103096. Shiyu Yang is supported by NSFC61802127 and Shanghai Sailing Program 18YF1406700.

\bibliographystyle{IEEEtran}
{
\small
\bibliography{main}

\begin{thebibliography}{10}
\providecommand{\url}[1]{#1}
\csname url@samestyle\endcsname
\providecommand{\newblock}{\relax}
\providecommand{\bibinfo}[2]{#2}
\providecommand{\BIBentrySTDinterwordspacing}{\spaceskip=0pt\relax}
\providecommand{\BIBentryALTinterwordstretchfactor}{4}
\providecommand{\BIBentryALTinterwordspacing}{\spaceskip=\fontdimen2\font plus
\BIBentryALTinterwordstretchfactor\fontdimen3\font minus
  \fontdimen4\font\relax}
\providecommand{\BIBforeignlanguage}[2]{{%
\expandafter\ifx\csname l@#1\endcsname\relax
\typeout{** WARNING: IEEEtran.bst: No hyphenation pattern has been}%
\typeout{** loaded for the language `#1'. Using the pattern for}%
\typeout{** the default language instead.}%
\else
\language=\csname l@#1\endcsname
\fi
#2}}
\providecommand{\BIBdecl}{\relax}
\BIBdecl

\bibitem{peng2018efficient}
Y.~Peng, Y.~Zhang, W.~Zhang, X.~Lin, and L.~Qin, ``Efficient probabilistic
  k-core computation on uncertain graphs,'' in \emph{2018 IEEE 34th
  International Conference on Data Engineering (ICDE)}.\hskip 1em plus 0.5em
  minus 0.4em\relax IEEE, 2018, pp. 1192--1203.

\bibitem{DBLP:journals/vldb/LiuYLQZZ20}
B.~Liu, L.~Yuan, X.~Lin, L.~Qin, W.~Zhang, and J.~Zhou, ``Efficient
  ({\(\alpha\)}, {\(\beta\)})-core computation in bipartite graphs,''
  \emph{{VLDB} J.}, vol.~29, no.~5, pp. 1075--1099, 2020.

\bibitem{peng2020answering}
Y.~Peng, Y.~Zhang, X.~Lin, L.~Qin, and W.~Zhang, ``Answering billion-scale
  label-constrained reachability queries within microsecond,''
  \emph{Proceedings of the VLDB Endowment}, vol.~13, no.~6, pp. 812--825, 2020.

\bibitem{lai2019distributed}
L.~Lai, Z.~Qing, Z.~Yang, X.~Jin, Z.~Lai, R.~Wang, K.~Hao, X.~Lin, L.~Qin, W.~Zhang~\emph{et al.}, 
  ``Distributed subgraph matching on timely dataflow,'' in \emph{Proceedings of the VLDB Endowment}, 
   vol.~12, no.~10, pp. 1099--1112, 2019.

\bibitem{qin2019towards}
Y.~Peng, Y.~Zhang, X.~Lin, W.~Zhang, L.~Qin, and J.~Zhou, ``Towards bridging
  theory and practice: hop-constrained st simple path enumeration,'' in
  \emph{International Conference on Very Large Data Bases}.\hskip 1em plus
  0.5em minus 0.4em\relax VLDB Endowment, 2019.

\bibitem{DBLP:journals/pvldb/QiuCQPZLZ18}
X.~Qiu, W.~Cen, Z.~Qian, Y.~Peng, Y.~Zhang, X.~Lin, and J.~Zhou, ``Real-time
  constrained cycle detection in large dynamic graphs,'' \emph{{PVLDB}},
  vol.~11, no.~12, pp. 1876--1888, 2018.

\bibitem{chung2003eigenvalues}
F.~Chung, L.~Lu, and V.~Vu, ``Eigenvalues of random power law graphs,''
  \emph{Annals of Combinatorics}, vol.~7, no.~1, pp. 21--33, 2003.

\bibitem{besta2019graph}
M.~Besta, D.~Stanojevic, J.~D.~F. Licht, T.~Ben-Nun, and T.~Hoefler, ``Graph
  processing on fpgas: Taxonomy, survey, challenges,'' \emph{arXiv preprint
  arXiv:1903.06697}, 2019.

\bibitem{FPGA}
{Xilinx}, ``https://www.xilinx.com/products/boards-and-kits/alveo.html.''

\bibitem{ICWCMMC2007}
D.~Yue, X.~Wu, Y.~Wang, Y.~Li, and C.~Chu, ``A review of data mining-based
  financial fraud detection research,'' in \emph{International Conference on
  Wireless Communications, Networking and Mobile Computing}, 10 2007, pp. 5519
  -- 5522.

\bibitem{kimura2006tractable}
M.~Kimura and K.~Saito, ``Tractable models for information diffusion in social
  networks,'' in \emph{European conference on principles of data mining and
  knowledge discovery}.\hskip 1em plus 0.5em minus 0.4em\relax Springer, 2006,
  pp. 259--271.

\bibitem{article_leser}
U.~Leser, ``A query language for biological networks,'' \emph{Bioinformatics},
  vol.~21, pp. ii33--9, 10 2005.

\bibitem{DBLP:conf/iwoca/RizziSS14}
R.~Rizzi, G.~Sacomoto, and M.~Sagot, ``Efficiently listing bounded length
  st-paths,'' in \emph{IWOCA}, 2014, pp. 318--329.

\bibitem{grossi2018efficient}
R.~Grossi, A.~Marino, and L.~Versari, ``Efficient algorithms for listing k
  disjoint st-paths in graphs,'' in \emph{Latin American Symposium on
  Theoretical Informatics}.\hskip 1em plus 0.5em minus 0.4em\relax Springer,
  2018, pp. 544--557.

\bibitem{bohmova2018computing}
K.~B{\"o}hmov{\'a}, L.~H{\"a}fliger, M.~Mihal{\'a}k, T.~Pr{\"o}ger,
  G.~Sacomoto, and M.-F. Sagot, ``Computing and listing st-paths in public
  transportation networks,'' \emph{Theory of Computing Systems}, vol.~62,
  no.~3, pp. 600--621, 2018.

\bibitem{bookknuth11}
D.~E. Knuth, \emph{The Art of Computer Programming, Volume 4A: Combinatorial
  Algorithms}.\hskip 1em plus 0.5em minus 0.4em\relax Addison-Wesley
  Professional, 2011.

\bibitem{DBLP:conf/ambn/YasudaSM17}
N.~Yasuda, T.~Sugaya, and S.~Minato, ``Fast compilation of s-t paths on a graph
  for counting and enumeration,'' in \emph{Proceedings of the 3rd Workshop on
  Advanced Methodologies for Bayesian Networks, {AMBN}}, 2017, pp. 129--140.

\bibitem{DBLP:conf/soda/BirmeleFGMPRS13}
E.~Birmel{\'{e}}, R.~A. Ferreira, R.~Grossi, A.~Marino, N.~Pisanti, R.~Rizzi,
  and G.~Sacomoto, ``Optimal listing of cycles and st-paths in undirected
  graphs,'' in \emph{SODA}, 2013, pp. 1884--1896.

\bibitem{DBLP:journals/ipl/Golovko72}
G.~L. D. and K.~N. P., ``Identifying certain types of parts of a graph and
  computing their number,'' \emph{Ukrainian Mathematical Journal}, vol.~24,
  no.~3, pp. 313--321, 1972.

\bibitem{DBLP:journals/corr/GiscardKW16}
P.~Giscard, N.~Kriege, and R.~C. Wilson, ``A general purpose algorithm for
  counting simple cycles and simple paths of any length,'' \emph{CoRR}, vol.
  abs/1612.05531, 2016.

\bibitem{gotthilf2009improved}
Z.~Gotthilf and M.~Lewenstein, ``Improved algorithms for the k simple shortest
  paths and the replacement paths problems,'' \emph{Information Processing
  Letters}, vol. 109, no.~7, pp. 352--355, 2009.

\bibitem{DBLP:journals/Yen71}
J.~Y. Yen, ``Finding the k shortest loopless paths in a network,''
  \emph{Management Science}, vol.~17, no.~11, pp. 712--716, 1971.

\bibitem{rivera2016mathematical}
J.~C. Rivera, H.~M. Afsar, and C.~Prins, ``Mathematical formulations and exact
  algorithm for the multitrip cumulative capacitated single-vehicle routing
  problem,'' \emph{European Journal of Operational Research}, vol. 249, no.~1,
  pp. 93--104, 2016.

\bibitem{shi2017multi}
N.~Shi, S.~Zhou, F.~Wang, Y.~Tao, and L.~Liu, ``The multi-criteria constrained
  shortest path problem,'' \emph{Transportation Research Part E: Logistics and
  Transportation Review}, vol. 101, pp. 13--29, 2017.

\bibitem{liu2017finding}
H.~Liu, C.~Jin, B.~Yang, and A.~Zhou, ``Finding top-k shortest paths with
  diversity,'' \emph{IEEE Transactions on Knowledge and Data Engineering},
  vol.~30, no.~3, pp. 488--502, 2017.

\bibitem{talarico2015k}
L.~Talarico, K.~S{\"o}rensen, and J.~Springael, ``The k-dissimilar vehicle
  routing problem,'' \emph{European Journal of Operational Research}, vol. 244,
  no.~1, pp. 129--140, 2015.

\bibitem{tommiska2001dijkstra}
M.~Tommiska and J.~Skytt{\"a}, ``Dijkstra’s shortest path routing algorithm
  in reconfigurable hardware,'' in \emph{International Conference on Field
  Programmable Logic and Applications}.\hskip 1em plus 0.5em minus 0.4em\relax
  Springer, 2001, pp. 653--657.

\bibitem{zhou2015accelerating}
S.~Zhou, C.~Chelmis, and V.~K. Prasanna, ``Accelerating large-scale
  single-source shortest path on fpga,'' in \emph{2015 IEEE International
  Parallel and Distributed Processing Symposium Workshop}.\hskip 1em plus 0.5em
  minus 0.4em\relax IEEE, 2015, pp. 129--136.

\bibitem{bondhugula2006parallel}
U.~Bondhugula, A.~Devulapalli, J.~Fernando, P.~Wyckoff, and P.~Sadayappan,
  ``Parallel fpga-based all-pairs shortest-paths in a directed graph,'' in
  \emph{Proceedings 20th IEEE International Parallel \& Distributed Processing
  Symposium}.\hskip 1em plus 0.5em minus 0.4em\relax IEEE, 2006, pp. 10--pp.

\bibitem{betkaoui2012parallel}
B.~Betkaoui, Y.~Wang, D.~B. Thomas, and W.~Luk, ``Parallel fpga-based all pairs
  shortest paths for sparse networks: A human brain connectome case study,'' in
  \emph{22nd International Conference on Field Programmable Logic and
  Applications (FPL)}.\hskip 1em plus 0.5em minus 0.4em\relax IEEE, 2012, pp.
  99--104.

\bibitem{CSR}
{Compressed sparse row}, ``https://en.wikipedia.org/wiki/sparse\_matrix.''

\end{thebibliography}
}
\end{document}